\documentclass[fleqn,usenatbib]{mnras}

\usepackage{newtxtext,newtxmath}

\usepackage[T1]{fontenc}

\DeclareRobustCommand{\VAN}[3]{#2}
\let\VANthebibliography\thebibliography
\def\thebibliography{\DeclareRobustCommand{\VAN}[3]{##3}\VANthebibliography}

\usepackage{amsmath}	
\usepackage{comment}
\usepackage{graphicx}	
\usepackage[version=4]{mhchem}
\usepackage{threeparttable} 
\usepackage{xspace}

\begingroup
\catcode`\_=\active
\gdef_#1{\ensuremath{\sb{\mathrm{#1}}}}
\endgroup
\mathcode`\_=\string"8000
\catcode`\_=12

\newcommand{\hm}{H$_2$\xspace} 
\newcommand{\hi}{H~I\xspace} 
\newcommand{\hii}{H~II\xspace} 
\newcommand{\hei}{He~I\xspace} 
\newcommand{\heii}{He~II\xspace} 
\newcommand{\hp}{h_{P}} 

\newcommand{\intj}{{\rm J}} 

\newcommand{\nh}{n_{H}} 
\newcommand{\npop}{n_{pop3}\xspace} 
\newcommand{\npoptwo}{n_{pop2}\xspace} 

\newcommand{\xe}{x_{e}} 

\newcommand{\hcc}{H~cm$^{-3}$\xspace} 
\newcommand{\mpch}{{\rm Mpc}/h} 
\newcommand{\msun}{{\rm M}_{\odot}} 
\newcommand{\myr}{{\rm Myr}}    
\newcommand{\pc}{{\rm pc}} 
\newcommand{\pch}{{\rm pc}/h} 
\newcommand{\zsun}{{\rm Z}_{\odot}}

\newcommand{\dd}{\mathrm{d}}

\newcommand{\colossus}{{\sc colossus}\xspace} 
\newcommand{\music}{{\sc music}\xspace} 
\newcommand{\ramses}{{\sc ramses}\xspace} 
\newcommand{\ramsesrt}{{\sc ramses-rt}\xspace} 
\newcommand{\rockstar}{{\sc rockstar}\xspace} 

\newcommand{\eg}{e.g.}

\title[First stars in an X-ray background]{Population~III star formation in an X-ray background: IV.\\ On-the-fly calculation of radiation backgrounds and their impact on the intergalactic medium}

\author[J. Park and M. Ricotti]{
Jongwon Park$^{1}$\thanks{E-mail: jwpark5064@yonsei.ac.kr}
and Massimo Ricotti$^{2}$
\\
$^{1}$Department of Astronomy, Yonsei University, Seoul 03722, Republic of Korea\\
$^{2}$Department of Astronomy, University of Maryland, College Park, MD 20742, USA\\
}

\date{Accepted XXX. Received YYY; in original form ZZZ}

\pubyear{\the\year{}}

\begin{document}
\label{firstpage}
\pagerange{\pageref{firstpage}--\pageref{lastpage}}
\maketitle

\begin{abstract}
In this paper, part of a series on the effects of X-ray sources in promoting Population~III (Pop~III) star formation, we investigate the ionisation and heating of the intergalactic medium (IGM) and the consequent enhancement of molecular hydrogen (\hm) and Pop~III formation using cosmological zoom-in simulations. We adopt a X-ray feedback model in which X-rays originate solely from Pop~III supernovae, and compute the global X-ray and Lyman-Werner (LW) radiation backgrounds on-the-fly during the simulation of a mean-density region of the Universe. This approach self-consistently captures the feedback loop between Pop~III stars and the radiation backgrounds they produce. Pop~III supernovae generate a weak X-ray background ($\intj_{X,21} \sim 10^{-5}$) and a moderate LW background ($\intj_{LW,21} \sim 10^{-1}$); the latter intensifies below $z \approx 12$ ($\intj_{LW,21} \sim 10^{1}$--$10^{2}$) with the onset of Pop~II star formation. Applying these backgrounds to regions of varying mean density produces a net positive X-ray feedback that increases the Pop~III number density, with stronger enhancement in underdense regions. The positive feedback is more pronounced when the X-ray background is computed on-the-fly rather than by post-processing, demonstrating the importance of the feedback loop. The X-ray background also raises the Thomson scattering optical depth at high redshift, while the total optical depth remains consistent with Planck~2018 constraints. Because our model includes only Pop~III supernovae as X-ray sources, stronger X-ray feedback is expected when additional sources are included, as will be explored in future work.
\end{abstract}





\section{Introduction}

Radiation backgrounds in the early Universe play a crucial role in the formation of the first stars, galaxies, and black holes, and carry important cosmological implications. Two types of radiation background have received particular attention: far-UV (FUV) radiation in the Lyman-Werner bands (LW, $11.2~{\rm eV} \leq E \leq 13.6~{\rm eV}$), and X-rays ($E \gtrsim 0.2~{\rm keV}$). X-ray radiation at high redshifts ($z \gtrsim 10$) has been studied for its potential influence on Pop~III star formation (\citealp{oh_reionization_2001}; \citealp*{venkatesan_heating_2001}; \citealp*{machacek_effects_2003}; \citealp{jeon_radiative_2014,hummel_first_2015,ricotti_x-ray_2016}; \citealp*{park_population_2021a}) and cosmic reionisation (\citealp*{haiman_radiative_2000}; \citealp{glover_radiative_2003,ricotti_x-ray_2004}; \citealp*{ricotti_x-ray_2005}). Recent \textit{JWST} observations have revealed a significant population of high-redshift black holes \citep[e.g., AGN and little red dots (LRDs),][]{juodzbalis_epochs_2023,matthee_little_2024,ananna_x-ray_2024,yue_stacking_2024,fujimoto_uncover_2024}, further underscoring the importance of X-ray radiation in the early Universe.

Modelling the LW and X-ray backgrounds self-consistently in cosmological simulations is challenging because of their long mean free paths. High spatial resolution is also needed to resolve the first X-ray sources — the remnants of Pop~III stars. Such sources must exist unless all Pop~III stars collapse directly into black holes without exploding, an unlikely outcome for plausible Pop~III mass functions. At the same time, a sufficiently large volume is required to capture the contribution of distant sources to the global background. Such efforts have already been made by some authors by attaching several simulation volumes to model LW backgrounds \citep{ahn_inhomogeneous_2009,incatasciato_modelling_2023}.

This paper describes a new method for self-consistently including LW and X-ray radiation backgrounds in small-volume, high-resolution cosmological simulations, and examines their impact on the IGM. The implications for Pop~III star formation are presented in the companion paper \citep[hereafter PR26b]{park_population_2026}. We first review the literature on X-ray feedback on star formation and discuss how estimates of the early X-ray background can be improved.

\subsection{X-ray Effects on Pop~III Star Formation}
\label{sec:intro_pop3}
The importance of X-rays stems from the opacity of the early IGM to hydrogen-ionising UV (EUV) radiation. EUV photons from Pop~III stars are rapidly absorbed by the nearly neutral medium ($\xe \sim 10^{-4}$). X-ray photons produced by Pop~III supernovae, high-mass X-ray binaries, and AGN, by contrast, can penetrate this neutral IGM \citep[see fig.~1 in][hereafter R16]{ricotti_x-ray_2016}, building up a global background. This background ionises both the diffuse IGM and gas within minihaloes ($M_{halo} \sim 10^6~\msun$), enhancing the gas-phase formation of \hm which, in the absence of dust, relies on the catalyst H$^-$:
\begin{equation}
    \label{eq:hminus}
~~~~~~~~\ce{ H + e^- -> H^- + \gamma }, ~~~~
        \ce{ H + H^- -> H_2 + e^- },
\end{equation}
as suggested by several authors (\citealp*{haiman_radiative_2000}; \citealp{oh_reionization_2001, glover_radiative_2003}; R16). X-ray photons are particularly efficient at ionising a nearly neutral medium: energetic photoelectrons trigger multiple secondary ionisations while depositing only a small fraction of their energy as heat \citep{shull_x-ray_1985, ricotti_fate_2002a}. For example, in a neutral medium ($\xe \sim 10^{-4}$), a photon with $E \sim 200~{\rm eV}$ can secondarily ionise $\sim 20$ hydrogen atoms, with only $\sim 10~\%$ of its energy converted to thermal energy \citep{park_population_2021a}.

Enhanced \hm formation (equation~(\ref{eq:hminus})) boosts \hm cooling, promoting Pop~III star formation and shaping stellar properties. In a series of papers \citep*[hereafter Papers~I--IV]{park_population_2021a, park_population_2021b, park_population_2023, park_origin_2024}, we found that an X-ray background leads to a less massive quasi-hydrostatic core due to enhanced \hm cooling, resulting in lower Pop~III stellar masses, reduced multiplicity, a steeper initial mass function (IMF), and more compact Pop~III binaries. In Paper~I, we also examined how X-rays modify the conditions required for Pop~III star formation.

The conditions for Pop~III star formation have been studied by many authors, with particular focus on the critical halo mass (\citealp*{machacek_simulations_2001}; \citealp{machacek_effects_2003}; R16; \citealp{schauer_influence_2019, skinner_cradles_2020, schauer_influence_2021}; \citealp*{kulkarni_critical_2021,nebrin_starbursts_2023}; \citealp{bovill_kindling_2024, hirano_formation_2025}). Although its exact definition varies, the core idea is that inefficient \hm cooling or IGM heating prevents Pop~III star formation in minihaloes below a critical mass. A moderate X-ray background can overcome this barrier by enhancing \hm formation and cooling, and by compensating for the negative LW feedback, provided that the associated increase in the IGM Jeans mass from X-ray heating remains sub-dominant (\citealp{haiman_radiative_2000, glover_radiative_2003}; R16; Paper~I). The resulting reduction in the critical mass increases the Pop~III number density ($\npop$, the number of Pop~III stars per comoving volume $(\mpch)^3$), with potential implications for the PISNe rate, IGM metal enrichment, and first-galaxy formation.

\subsection{X-ray Effects on Reionisation}
\label{sec:intro_re}

There is broad consensus that the neutral IGM was completely reionised by $z \sim 6$ \citep{fan_constraining_2006}. The timing of reionisation is closely tied to the Thomson scattering optical depth $\tau_{e}$: a higher value indicates the Universe remained ionised for longer. Identifying the dominant ionising sources — EUV and X-rays — and assessing their relative contributions during reionisation are therefore key questions in cosmology.

UV radiation from dwarf galaxies ($10^8~\msun \lesssim M_{halo} \lesssim 10^{11}~\msun$) is widely accepted as the dominant driver of reionisation \citep{haehnelt_ionizing_2001, kimm_feedback-regulated_2017, madau_radiation_2017}, but the contribution of X-rays has long been debated (\citealp{oh_reionization_2001, machacek_effects_2003, glover_radiative_2003, ricotti_x-ray_2004}; \citealp*{ricotti_x-ray_2005}; \citealp{ahn_detecting_2012, xu_x-ray_2016}). Early models invoked X-ray pre-ionisation to explain the high optical depth reported by the first-year \textit{WMAP} results \citep[$\tau_{e} = 0.17$,][]{bennett_first-year_2003}, but subsequent \textit{WMAP} and \textit{Planck} measurements revised $\tau_{e}$ downward \citep{bennett_nine-year_2013, planck_collaboration_planck_2020}. The most recent estimate \citep[$\tau_{e} = 0.054$,][]{planck_collaboration_planck_2020} is nearly three times lower than the original \textit{WMAP} value, apparently limiting the role of an early X-ray background. Nevertheless, recent observations suggest a non-negligible pre-reionisation X-ray background. HERA results indicate substantial X-ray heating at $z \sim 8$ \citep{abdurashidova_hera_2022}, and \textit{JWST} observations point to a high incidence of high-redshift AGN \citep{juodzbalis_epochs_2023,fujimoto_uncover_2024}. Models of the recently discovered LRD population at $z \sim 5$ suggest they are powered by massive black holes embedded in dense gas \citep{matthee_little_2024,rusakov_jwsts_2025,kido_black_2025}; while LRDs are not strong X-ray emitters at this stage of their evolution, their discovery provides indirect evidence for a large population of high-redshift black holes that may have contributed to an early X-ray background at $z > 6$. Together, these clues motivate a reassessment of the X-ray background's impact on the IGM and reionisation.


\begin{figure}
    \includegraphics[width=\columnwidth]{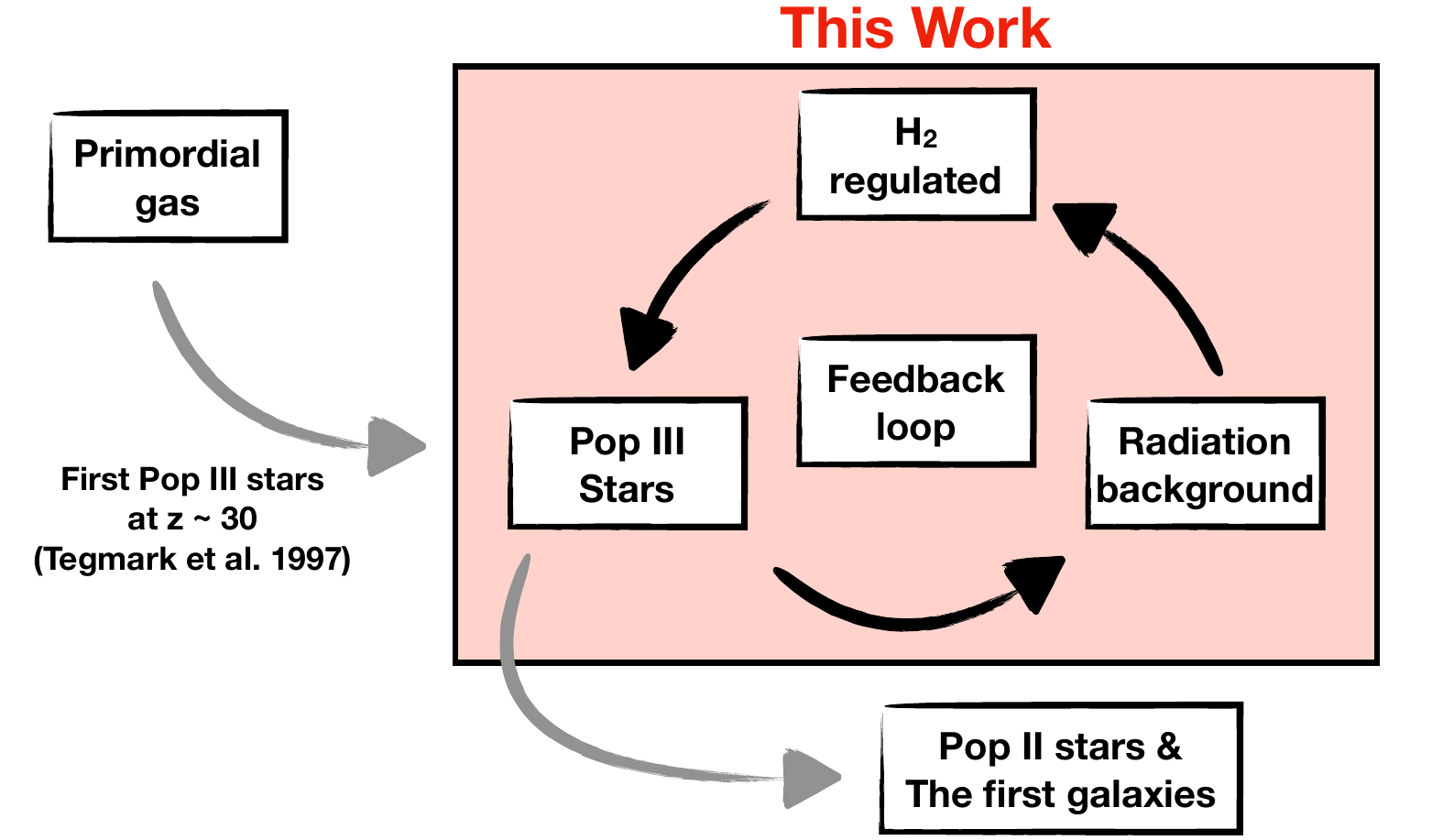}
    \caption{Schematic diagram illustrating the relationship between Pop~III stars and radiation background. The scope of this paper is highlighted with a bright red box. Since the formation of the first Pop~III stars at $z \sim 30$ \citep{tegmark_how_1997}, Pop~III stars and the radiation background interact to form a feedback loop. We briefly comment on the potential influence of the background on the formation of the first galaxies in PR26b but this will be handled in detail in future work.}
    \label{fig:loop}
\end{figure}

\subsection{Improvements on Previous Works and Motivation}
\label{sec:motivation}

Assessing the role of an X-ray background in Pop~III star formation and reionisation requires improvements in the methods used to estimate its intensity. Paper~I took a first step toward understanding the effects of external radiation backgrounds on Pop~III star formation, but had several limitations.
\begin{enumerate}
    \item Paper~I explored X-ray and LW radiation using grids of zoom-in simulations of dark matter minihaloes at $z > 9$, spanning 7 LW intensities ($0$ to $10^2~{\mathrm J}_{21}$\footnote{${\mathrm J}_{21} = 10^{-21}~{\rm erg}~{\rm s}^{-1}~{\rm cm}^{-2}~{\rm Hz}^{-1}~{\rm sr}^{-1}$}) and 7 X-ray intensities ($0$ to $10^{-1}~{\mathrm J}_{21}$), similar to the pioneering studies of \citet{machacek_simulations_2001} and \citet{machacek_effects_2003} who varied the normalisation of the radiation backgrounds and systematically investigated their impacts on Pop~III star formation. Which feedback mechanism dominates (X-ray ionisation, X-ray heating, or LW photodissociation) depends on the radiation field intensities, which must be estimated within a cosmological context because a feedback loop couples the radiation backgrounds to the number and masses of Pop~III stars, and vice versa (see R16)).
    
    \item Each simulation zoomed in on a single isolated halo. However, many cosmological simulations have shown that star formation can be suppressed by supernova feedback \citep{wise_resolving_2008, jeon_recovery_2014} and by radiative feedback from stars in neighbouring haloes \citep{regan_emergence_2020}. Both local and global radiation backgrounds must therefore be accounted for.
\end{enumerate}

Building on pioneering estimates of the early X-ray background \citep{haardt_radiative_2012,jeon_radiative_2014,xu_heating_2014,ahn_spatially_2015,hummel_first_2015,xu_x-ray_2016,madau_radiation_2017}, we identify three aspects that need to be improved.
\begin{enumerate}
    \item The calculation should cover star formation at high redshifts ($z \gtrsim 15$), when Pop~III stars are active. Previous estimates \citep{haardt_radiative_2012, madau_radiation_2017} did not include Pop~III stars and their remnants at these early epochs.

    \item The feedback loop must be included self-consistently. As illustrated in Fig.~\ref{fig:loop}, an X-ray background enhances Pop~III star formation, which in turn produces supernovae and black holes that emit further X-rays. Several studies \citep{xu_heating_2014, ahn_spatially_2015, xu_x-ray_2016} estimated the X-ray background by post-processing the \textit{Renaissance} simulations \citep{xu_population_2013}, and therefore did not capture this real-time interaction.

    \item The contribution of distant sources must be included. X-ray photons have mean free paths ($\gtrsim 10~\mpch$) far exceeding the box sizes ($\sim 1~\mpch$) needed to resolve individual Pop~III stars. Studies that transferred X-ray photons only within the simulation box \citep{jeon_first_2012, jeon_radiative_2014} may therefore have missed the contribution of distant sources.
\end{enumerate}
Here we implement a new on-the-fly method to calculate the X-ray and LW backgrounds that incorporates both the feedback loop and the contribution of distant Pop~III stars, and use it to investigate the impact of the X-ray background on the IGM and Pop~III star formation.

The paper is structured as follows. Section~\ref{sec:method} describes the cosmological simulations and the on-the-fly background calculation. Section~\ref{sec:xray} presents the impact of the global LW and X-ray backgrounds on primordial gas and shows how the on-the-fly approach affects Pop~III star counts. Section~\ref{sec:discussion} contains discussion and a summary of the results.

\begin{table}
    \caption{Photon energy groups.}
    \centering
    \label{tab:bin}
    \begin{threeparttable}
    	\begin{tabular}{ | c | c | c | l | }
		    \hline
            Group & $h_{P}\nu_1$~(eV) & $h_{P}\nu_2$~(eV) & Roles \\
		    \hline
            1 & 11.20 & 13.60 & \hm dissociation (LW) \\
            2 & 13.60 & 15.20 & \hi ionisation \\
            3 & 15.20 & 24.60 & \hi, \hm ionisation \\
            4 & 24.60 & 54.42 & \hi, \hm, \hei ionisation \\
            5 & 54.42 & 200 & \hi, \hm, \hei, \heii ionisation \\
            6 & 200 & $\infty$ & \hi, \hm, \hei, \heii ionisation,\\
            & & & X-ray radiation (Pop~III SN) \\
            \hline
	    \end{tabular}
    \end{threeparttable}
\end{table}


\begin{figure}
    \centering
	\includegraphics[width=0.48\textwidth]{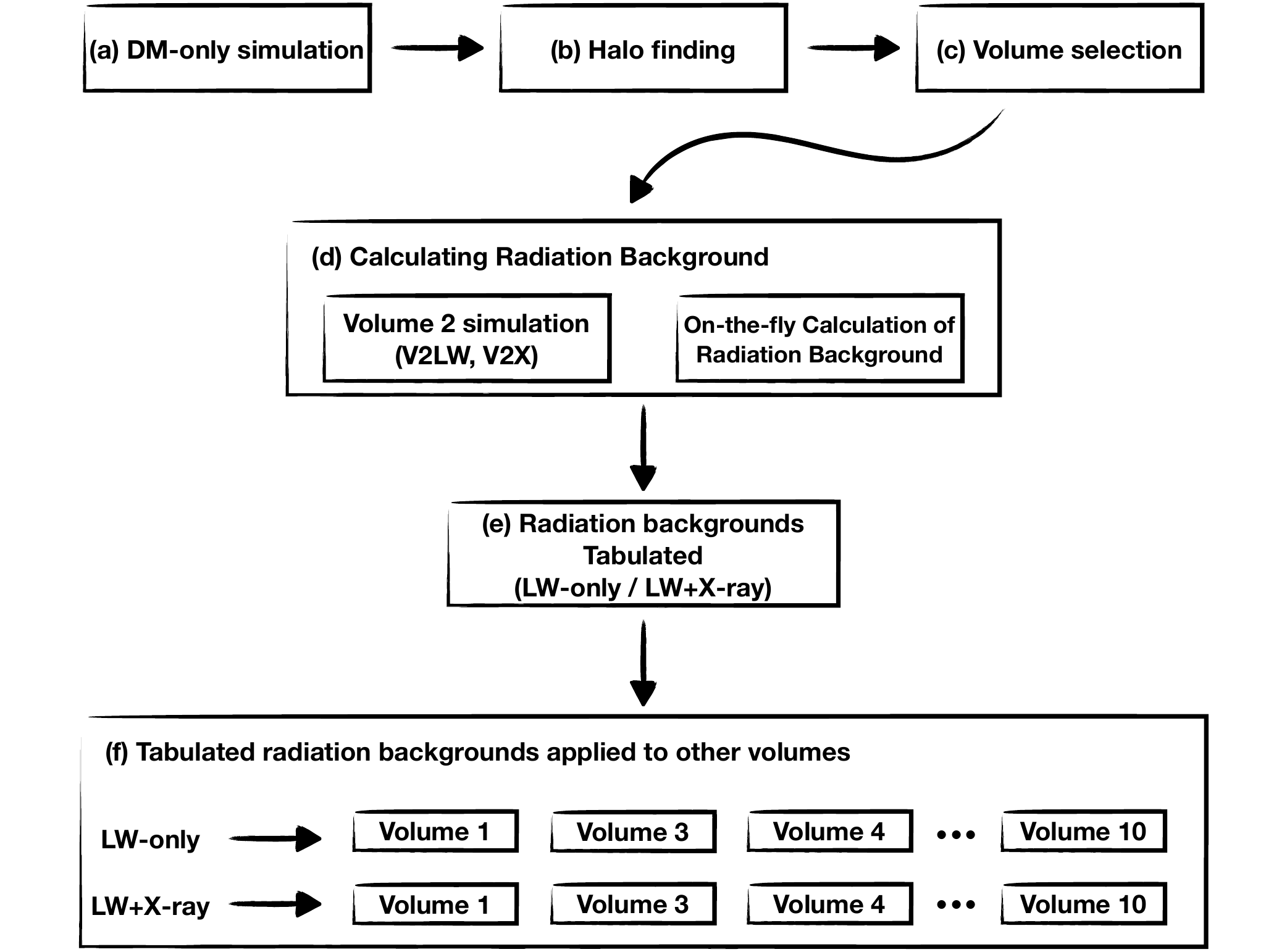}
    \caption{Schematic diagram of the work process. \textbf{(a}) We performed two dark matter-only simulations. \textbf{(b)} DM haloes were identified using \rockstar halo finder \citep{behroozi_rockstar_2013}. \textbf{(c)} Based on the halo properties (masses and positions), we selected sub-volumes. \textbf{(d)} Volume 2 was designated as the representative volume, for which we carried out zoom-in simulations to calculate the radiation backgrounds on-the-fly. LW-only simulation (V2LW) and LW + X-ray simulation (V2X) were performed for this purpose. \textbf{(e)} Each resulting radiation background (spectrum) was tabulated as a function of redshift. \textbf{(f)} These LW-only and LW + X-ray backgrounds were applied to the other sub-volumes. For instance, we ran two simulations for Volume 1 (V1LW and V1X).}
    \label{fig:order}
\end{figure}

\begin{table*}
    \caption{Summary of the simulations}
    \centering
    \footnotesize
    \begin{threeparttable}
        \begin{tabular}{ | l | l | c | c | c | c | c | c | c | c | l | c | c | c | c | }
		\hline
            ~~Title & ~~Volume & $L_{box}$\tnote{a} & $L_{zoom}$\tnote{b} & $L_{tar}$\tnote{c} & $l_{min}$\tnote{d} & $l_{max}$\tnote{e} & $\Delta x_{com}$\tnote{f} & $\Delta x_{phy}$\tnote{g} & $m_{DM}\tnote{h}$ & ~~Radiation\tnote{i} & $N_{\rm Pop3,fd}$\tnote{j} & $z_{final}$ &
            $\delta+1$\tnote{k} \\
            
		\hline
            DM8\tnote{\textdagger} & Entire & $8$ &  & & $9$ & $14$ & $488.3$ & $72.4$ & $2.813 \times 10^5$ & & & $9$ & \\ 
            
		\hline
            DM4\tnote{\textdagger} & Entire & $4$ &  & & $8$ & $13$ & $488.3$ & $72.4$ & $2.813 \times 10^5$ & & & $9$ &  \\ 
            
            \hline
            V1LW & Volume 1 & $8$ & $1.2$ & $1$ & $8$ & $21$ & $3.81$ & $0.57$ & $4,395$ & LW  & $272$ & $13.6$ & 1.99   \\
            
            \hline
            V1X & Volume 1 & $8$ & $1.2$ & $1$ & $8$ & $21$ & $3.81$ & $0.57$ & $4,395$  & LW + X-ray  & $332$ & $13.9$ & 1.95  \\
            
            \hline
            V2LW & Volume 2 & $8$ & $1.2$ & $1$ & $8$ & $21$ & $3.81$ & $0.57$ & $4,395$ & LW  & $92$ & $9$ & 1.18 \\
            
            \hline
            V2X & Volume 2 & $8$ & $1.2$ & $1$ & $8$ & $21$ & $3.81$ & $0.57$ & $4,395$ & LW + X-ray  & $165$ & $9$ & 1.18 \\
            
            \hline
            V3LW & Volume 3 & $8$ & $1.2$ & $1$ & $8$ & $21$ & $3.81$ & $0.57$ & $4,395$ & LW  & $1$ & $9$ & 0.66 \\
            
            \hline
            V3X & Volume 3 & $8$ & $1.2$ & $1$ & $8$ & $21$ & $3.81$ & $0.57$ & $4,395$ & LW + X-ray  & $7$ & $9$ & 0.66 \\
            
            \hline
            V4LW & Volume 4 & $8$ & $1.2$ & $1$ & $8$ & $21$ & $3.81$ & $0.57$ & $4,395$ & LW  & $1$ & $9$ & 0.65 \\
            
            \hline
            V4X & Volume 4 & $8$ & $1.2$ & $1$ & $8$ & $21$ & $3.81$ & $0.57$ & $4,395$ & LW + X-ray  & $5$ & $9$ & 0.65 \\
            
            \hline
            V5LW & Volume 5 & $8$ & $1.2$ & $1$ & $8$ & $21$ & $3.81$ & $0.57$ & $4,395$ & LW  & $0$ & $9$ & 0.58 \\
            
            \hline
            V5X & Volume 5 & $8$ & $1.2$ & $1$ & $8$ & $21$ & $3.81$ & $0.57$ & $4,395$ & LW + X-ray  & $2$ & $9$ & 0.58 \\
            
            \hline
            V6LW & Volume 6 & $4$ & $0.6$ & $0.5$ & $8$ & $21$ & $1.90$ & $0.28$ & $549$ & LW  & $0$ & $9$ & 0.46 \\
            
            \hline
            V6X & Volume 6 & $4$ & $0.6$ & $0.5$ & $8$ & $21$ & $1.90$ & $0.28$ & $549$ & LW + X-ray  & $1$ & $9$ & 0.46 \\
            
            \hline
            V7LW & Volume 7 & $4$ & $0.6$ & $0.5$ & $8$ & $21$ & $1.90$ & $0.28$ & $549$ & LW  & $1$ & $9$ & 0.61 \\
            
            \hline
            V7X & Volume 7 & $4$ & $0.6$ & $0.5$ & $8$ & $21$ & $1.90$ & $0.28$ & $549$ & LW + X-ray  & $2$ & $9$ & 0.61 \\
            
            \hline
            V8LW & Volume 8 & $4$ & $0.6$ & $0.5$ & $8$ & $21$ & $1.90$ & $0.28$ & $549$ & LW  & $0$ & $9$ & 0.50 \\
            
            \hline
            V8X & Volume 8 & $4$ & $0.6$ & $0.5$ & $8$ & $21$ & $1.90$ & $0.28$ & $549$ & LW + X-ray  & $0$ & $9$ & 0.50 \\
            
            \hline
            V9LW & Volume 9 & $4$ & $0.6$ & $0.5$ & $8$ & $21$ & $1.90$ & $0.28$ & $549$ & LW  & $0$ & $9$ & 0.44 \\
            
            \hline
            V9X & Volume 9 & $4$ & $0.6$ & $0.5$ & $8$ & $21$ & $1.90$ & $0.28$ & $549$ & LW + X-ray  & $1$ & $9$ & 0.44 \\
            
            \hline
            V10LW & Volume 10 & $4$ & $0.6$ & $0.5$ & $8$ & $21$ & $1.90$ & $0.28$ & $549$ & LW  & $1$ & $9$ & 0.69 \\
            
            \hline
            V10X~ & Volume 10 & $4$ & $0.6$ & $0.5$ & $8$ & $21$ & $1.90$ & $0.28$ & $549$ & LW + X-ray  & $1$ & $9$ & 0.69 \\
            
            \hline

	\end{tabular}
        \begin{tablenotes}
	        \item[a] Size of the entire box ($\mpch$).
                
                \item[b] Size of the zoom-in region ($\mpch$). This includes the target region and the padding.
                
                \item[c] Size of the target region ($\mpch$).
                
                \item[d] Minimum AMR level.
                
                \item[e] Maximum AMR level.
                
                \item[f] Smallest cell size ($\pch$, comoving). 
                
                \item[g] Smallest cell size at $z=9$ ($\rm pc$, physical). 
                
                \item[h] Mass of a DM particle ($\msun/h$). For the zoom-in simulations, we show the effective mass resolution (mass within the zoom-in region).
                
                \item[i] Types of the radiation background considered in the simulation. Note that simulations titled `X' include `both' LW and X-ray effects.
                
                \item[j] Total number of Pop~III stars forming in the fiducial region by $z = 9$.

                \item[k] Dark matter overdensity of the target region. $\delta + 1 = \rho / \rho_{mean}$. \\

                \item[\textdagger] This is the DM-only simulation and was utilized for the selection of the subvolumes.
        \end{tablenotes}
    \end{threeparttable}
    \label{tab:simulation}
\end{table*}


\begin{figure*}
    \includegraphics[width=\textwidth]{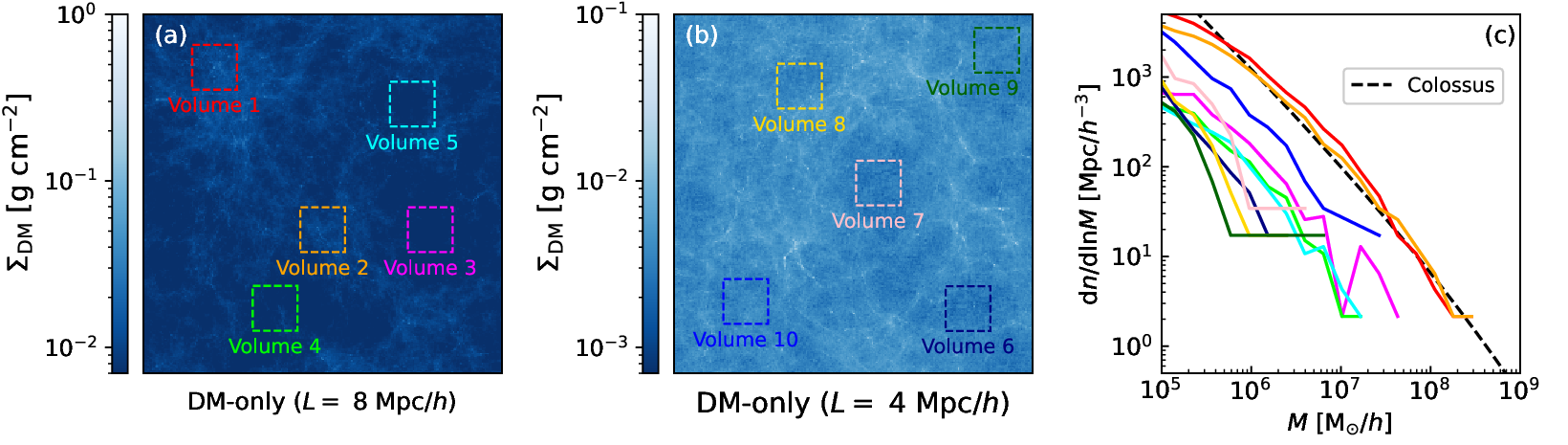}
    \caption{\textbf{Panels~a and b:} The projected DM density maps of the DM-only simulations of $L = 8$ and $4~\mpch$, respectively. Coloured boxes indicate the 10  sub-volumes chosen for zoom-in simulations. \textbf{Panel~c:} Halo mass functions of the zoom-in simulations of these sub-volumes. The mass functions at $z = 9$ are shown except for V1 (red, $z = 13.94$). Colours correspond to the sub-volumes marked in Panel~a and b. For reference, the PS mass function at $z = 9$, calculated using \colossus \citep{diemer_colossus_2018}, is shown as a dashed line. The mass function of Volume~2 (orange) is similar to the PS prediction; we therefore regard its Pop~III star formation history as representative of the cosmic average.}
    \label{fig:illust_dm}
\end{figure*}


\section{Simulations \& Methods}
\label{sec:method}

\subsection{Overview}
\label{sec:overview}

We performed radiative hydrodynamics (RHD) simulations using the adaptive mesh refinement (AMR) code \ramsesrt \citep{teyssier_cosmological_2002, rosdahl_ramses-rt_2013}. In the simulations, we distinguish ``local'' radiation background from ``global'' radiation backgrounds. The former originates from sources within the central zoom-in region, whereas the latter represents the cumulative radiation from virtual sources outside this region. The local background is divided into six photons groups according to their chemical roles, as summarised in Table~\ref{tab:bin}. Groups 1 and 6 correspond to ``local'' LW and X-ray radiation, respectively, while the remaining groups represent local ionising UV radiation. Photon densities in a cell are increased when stars or supernovae emit radiation and decreased when photons are absorbed by gas. These photons are transported using M1-closure implementation in \citep{rosdahl_ramses-rt_2013}. In contrast, global radiation is represented by a spatially uniform spectrum (Fig.~\ref{fig:global_spec}) and is not transported within the simulation using the M1-closure. We refer to the portions of this spectrum within the energy ranges 11.2--13.6~eV and $> 200$~eV as the global LW and global X-ray backgrounds, respectively. In summary, this work considers four key radiation components: local LW, local X-ray, global LW, and global X-rays.

Estimating these backgrounds across diverse environments requires accounting for mean-density, overdense, and underdense regions: at any given redshift, an underdense region has lower average star formation rates (SFRs) and a weaker local extragalactic background than an overdense region, but both share the same global background. To capture this diversity, we performed a suite of simulations following three steps.
\begin{enumerate}
    \item \textbf{Sub-volume selection.} We performed two dark matter-only simulations with box sizes of $L = 8~\mpch$ and $L = 4~\mpch$, selecting five sub-volumes from each. The sub-volumes have equal sizes ($1~\mpch$ and $0.5~\mpch$, respectively) but different numbers of dark matter haloes, representing diverse cosmic environments.
    
    \item \textbf{Global background calculation.} Using Volume~2 as the representative mean-density region, we ran two zoom-in simulations: one with LW radiation only and one with both LW and X-rays (with all X-ray physics disabled in the former). Radiation backgrounds were computed on-the-fly following Section~\ref{sec:onthefly} to capture the contribution of distant sources and the feedback loop between Pop~III stars and radiation. The resulting star formation rates were tabulated for use in the next step.
    
    \item \textbf{Simulations of remaining sub-volumes.} Assuming that the star formation rate density derived from Volume~2 is representative of the cosmic mean, we applied it to compute the global background in all other sub-volumes. For each, we ran two simulations: LW-only and LW + X-ray. In the LW-only simulation, both local and global X-ray radiation backgrounds remain zero.
\end{enumerate}
A schematic diagram of this workflow is provided in Fig.~\ref{fig:order}. We adopted cosmological parameters $h = 0.674$, $\Omega_{m} = 0.315$, $\Omega_{\Lambda} = 0.685$, $\Omega_{b} = 0.0493$, $\sigma_{8} = 0.811$, and $n_{s} = 0.965$ \citep{planck_collaboration_planck_2020}.

\subsection{Selection of Regions}
\label{sec:region}

We generated cosmological initial conditions for boxes of size $L_{box} = 8~\mpch^3$ and $4~\mpch^3$ using \music \citep{hahn_multi-scale_2011}, with a DM particle mass of $2.8 \times 10^5~\msun/h$. DM-only simulations (DM8 and DM4 in Table~\ref{tab:simulation}) were then run, and DM haloes identified at $z = 9$ using \rockstar \citep{behroozi_rockstar_2013}.

We sampled random sub-boxes of size $1~\mpch$ and $0.5~\mpch$ (from DM8 and DM4, respectively), computed their halo mass functions, and compared them with that of the full volume. Repeating this process, we selected five distinct sub-volumes from each parent simulation. An overview of the full boxes and selected sub-volumes is shown in Fig.~\ref{fig:illust_dm}.

Volume~1 (red) has a higher-than-average DM halo concentration and contains the most massive halo among all sub-volumes. Its RHD simulations were stopped at $z \sim 13.5/13.9$ (LW/LW+X-ray runs, respectively) due to their high computational cost; its halo mass function is shown at that redshift in Fig.~\ref{fig:illust_dm}. For all other volumes, mass functions at $z = 9$ are shown, alongside the Press-Schechter (PS) prediction \citep[dashed line,][]{press_formation_1974, sheth_large-scale_1999} from \colossus \citep{diemer_colossus_2018}.

Volume~2 (orange) has a mass function closely matching the PS prediction and represents a mean-density environment; we adopt it as the reference volume for computing radiation backgrounds (Section~\ref{sec:onthefly}). Volumes~3--10 correspond to underdense regions. The full simulation parameters, including DM overdensities, are listed in Table~\ref{tab:simulation}.

\subsection{Zoom-in Simulations}
\label{sec:zoom}

\begin{figure}
    \centering
	\includegraphics[width=0.48\textwidth]{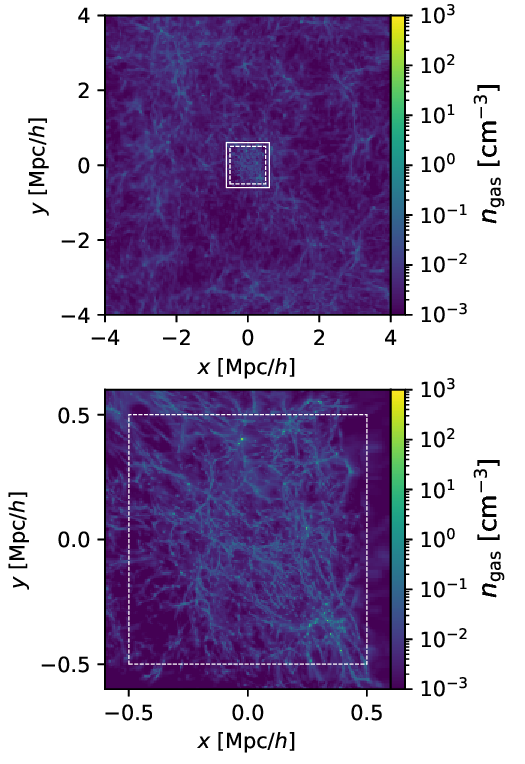}
    \caption{Snapshot of the zoom-in simulation V2X (see Table~\ref{tab:simulation}) at $z=9$. In the top panel, the gas density map of the full $8~\mpch$ box is depicted. Only gas cells within the central $1.2~\mpch$ zoom-in region, outlined by the solid white square, are eligible for refinement and star formation. The bottom panel provides an enlarged view of the zoom-in region. The dashed line marks the boundary between the target region (inner area) and the padding (outer area). All analyses in this work are based on the stars and gas located within the target region.}
    \label{fig:illust_zoom}
\end{figure}

To generate initial conditions for the zoom-in simulations, each selected sub-volume was relocated to the centre of the box. The minimum AMR level is set to 8, corresponding to $2^8 = 256$ base grid cells per side. Three additional levels of nested grids were created using \music within the base grid. The innermost nested grid covers $1.2~\mpch$ ($0.6~\mpch$ for Volumes\footnote{Values in parentheses throughout this subsection refer to Volumes~6--10, which are derived from DM4.}~6--10), with DM particles of mass $4{,}395~\msun/h$ ($549~\msun/h$), defining the effective DM mass resolution. Gas cells within the zoom-in region are refined when:
\begin{enumerate}
    \item a cell contains at least 8 DM particles; or
    \item the local Jeans length is smaller than $N_{J} = 4$ times the cell size \citep{truelove_jeans_1997}.
\end{enumerate}
Star formation is disabled outside the zoom-in region. Numerical artefacts near the boundaries due to differing DM particle masses are mitigated by designating a ``padding'' region around the boundary and excluding it from the analysis. The innermost $L_{tar} = 1~\mpch$ ($0.5~\mpch$) cube inside the padding is defined as the ``target'' region and contains all gas and particles used for the primary results. The ``zoom-in'' region refers to the target region together with its surrounding padding. An illustration is shown in Fig.~\ref{fig:illust_zoom}.


The maximum AMR level is 21, with the smallest cell size of $\Delta x = 3.81~{\rm pc}/h$ ($1.91~{\rm pc}/h$) in comoving units, or $\approx 0.57$~pc ($0.28$~pc) physical at $z=9$ (physical). The simulations employ the primordial chemical network, described in detail in Paper~I.

We stopped the zoom-in simulations of Volume~1 at $z \approx 14$ as they are cost-prohibitive. For the remaining volumes, we stopped the simulations at $z = 9$ motivated by the previous finding of \citet{jeon_first_2015} that star formation mode transitions at $z \sim 13$ although recent studies suggest Pop~III stars may continue to form down to $z \sim 6$ \citep{hegde_self-consistent_2023, Hegde2025Efficient}.

\subsection{Pop~III Stars}
\label{sec:pop3}

{\bf Star formation:} At each coarse time step, gas clumps are identified using the built-in halo finder \citep{bleuler_phew_2015}. A single Pop~III star particle with a mass of $M = 100~\msun$ is placed at the density peak of a clump if all of the following criteria are satisfied.
\begin{enumerate}
    \item The hydrogen number density at the peak ($\nh$) exceeds $10^4$~\hcc.

    \item The metallicity at the peak is below $10^{-4}~\zsun$, where $\zsun = 0.02$ is the default solar metallicity in \ramses.

    \item No other Pop~III stars are present within a radius of $100~\pc$.
\end{enumerate}
When a Pop~III star particle is created, $100~\msun$ is removed from all cells within the clump with $\nh > 10^3$~\hcc, in proportion to each cell's density. Although Pop~III stars span a wide mass range (\citealp{hirano_one_2014}; \citealp*{susa_mass_2014}; \citealp{hirano_primordial_2015, hosokawa_formation_2016, sugimura_birth_2020}), we adopt a fixed mass of $100~\msun$, consistent with the peak of the Pop~III IMF found in Papers~II and III. While Papers~I--III showed that X-ray feedback reduces both Pop~III mass and multiplicity, we assume here for simplicity that each halo forms a single $100~\msun$ Pop~III star unaffected by X-rays. Recent studies \citep{garcia_star_2023, sugimura_violent_2024} include Pop~III binarity in cosmological simulations; the effects of different IMFs on Pop~III and galaxy formation will be explored in future work.

Criterion (iii) has been adopted to limit the number of Pop~III star formation event to one per minihalo, unlike some earlier works which allow multiple Pop~III stars to form within the same halo \citep{skinner_cradles_2020,wells_connecting_2022}. It is well established that multiple Pop~III stars form through fragmentation of protostellar discs (\citealp*{stacy_first_2010}; \citealp{clark_formation_2011, sugimura_birth_2020}; Paper~I; Paper~III), a process that is not resolved in our cosmological simulations. Instead, we find that the limited spatial resolution can produce a numerical artefact in which the residual gas clump left behind after the formation of the first Pop~III star continues to accrete gas and subsequently forms additional Pop~III stars.\footnote{This process is distinct from the physically motivated Pop~III multiplicity described above. Primordial gas first forms a quasi-hydrostatic core with $n \sim 10^4$~\hcc, which then collapses under its self-gravity to form a protostellar disc. Multiple Pop~III stars are produced through disc fragmentation within the mass budget of this core (Paper~I, III). In contrast, the numerical artefact described here replenishes the core by accreting lower-density gas ($n < 10^4$~\hcc) within the halo.} This artificial multiple Pop~III star formation also differs from that of \citet{skinner_cradles_2020}, who allowed multiple Pop~III stars to form when multiple physically distinct high density cells ($n > 10^6$~cm$^{-3}$) were present within the same halo. Criterion (iii) suppresses this numerical artefact and maintains consistency with the global background calculation, which also assume a single Pop~III formation event per halo (Section~\ref{sec:onthefly}). As a side note, 100~pc is comparable to the Str\"{o}mgren radius of a $100~\msun$ Pop~III star in a medium of $n \sim 1$~\hcc and $T \sim 10^4$~K, and the virial radius of a $M \sim 10^{6}~\msun$ halo. The unresolved Pop~III multiplicity is reflected in PISN boost factor $\alpha_{HN}$ and is discussed below.


\begin{figure*}
    \centering
	\includegraphics[width=0.95\textwidth]{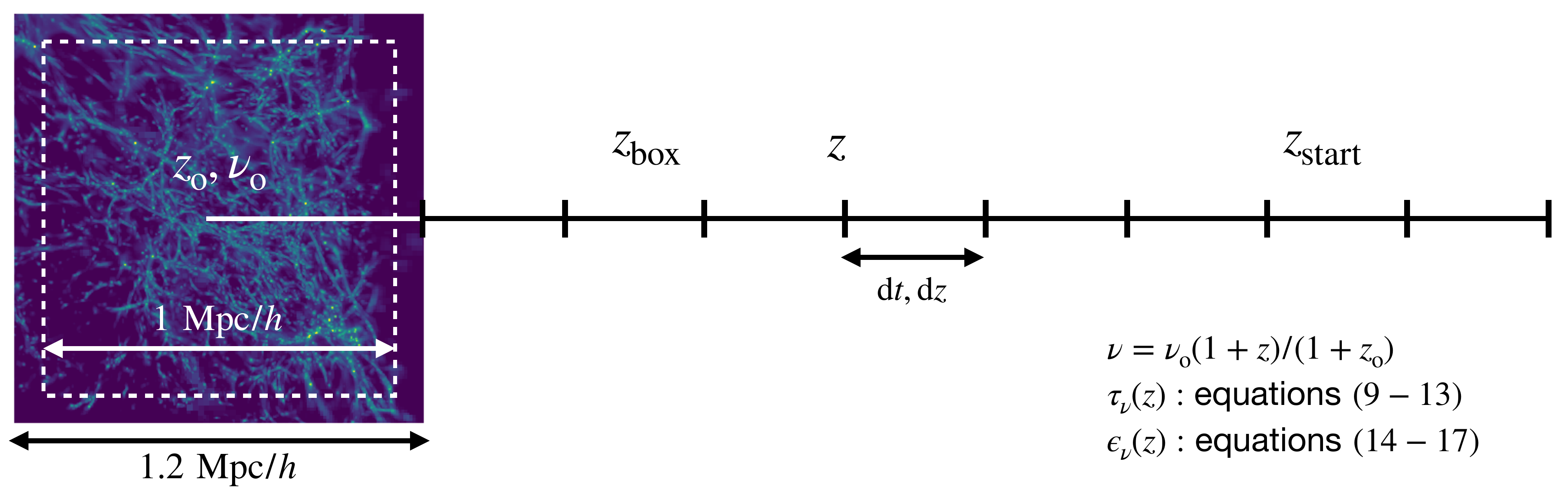}
    \caption{Schematic diagram of the on-the-fly background calculation. The $x$-axis represents both the distance and lookback time (redshift). We compute the specific intensity at the observed frequency $\nu_{o}$ in the zoom-in region at $z_{o}$ denoted by $\intj_{\nu_{o}}(z_{o})$, by numerically integrating equation~(\ref{eq:background}). The left-hand panel shows a snapshot including the target region (white dashed square) and the surrounding padding. We assume that sources existing between $z_{box}$ and $z_{start}$ contribute to the global radiation background. Photons emitted at frequency $\nu = \nu_{o} (1+z)/(1+z_{o})$ are redshifted and contribute to the intensity at the observed frequency $\nu_{o}$ at the current redshift $z_{o}$.}
    \label{fig:schematic}
\end{figure*}

\noindent {\bf Radiative feedback:} Once spawned, each Pop~III star particle injects UV photons into the $3 \times 3 \times 3$ cells centred on the star (27 cells) throughout its main-sequence lifetime \citep[$t_{pop3} = 2.77$~Myr for a $100~\msun$ star,][]{schaerer_properties_2002}. Following the results of \citet{schaerer_properties_2002}, we inject photons of Groups~1--5 (Table~\ref{tab:bin}). We assume the emission of Group~6 photons from a main-sequence star is negligible because even a $100~\msun$ star ($T \sim 10^5$~K) emits very little radiation in X-ray energies. Group~6 photons are instead emitted during the subsequent supernova remnant phase (see below). Injected photons are propagated using the M1-closure method \citep{rosdahl_ramses-rt_2013} and interact chemically with gas cells \citep{rosdahl_ramses-rt_2013,katz_interpreting_2017}.

\noindent {\bf Pop~III Supernova:} Each Pop~III star is assumed to explode as an energetic hypernova (HN) or pair-instability supernova (PISN) with energy $E_{HN} = 100 \times 10^{51}~{\rm erg} = 10^{53}$~erg, deposited uniformly across the 27 neighbouring cells. The star leaves no compact remnant and returns its full mass ($100~\msun$) to those cells, including $20~\msun$ of metals.

Some studies \citep[\eg,][]{wise_birth_2012, kimm_feedback-regulated_2017} draw Pop~III masses from a top-heavy IMF and apply mass-dependent stellar evolution. Following \citet{heger_nucleosynthetic_2002}, for instance, non-rotating Pop~III stars with $40~\msun < M < 140~\msun$ eject no material and collapse to black holes, while stars with $140~\msun < M < 260~\msun$ explode as PISNe leaving no BH remnant. Because the IMF, metal yield, and remnant mass of Pop~III stars are highly uncertain — depending on mass loss, explosion mechanism, rotation, and other factors — we adopt fixed values ($M = 100~\msun$, metal yield $= 20~\msun$, $E = 10^{53}$~erg) for simplicity.

According to the widely used stellar evolution of model of \citet{heger_nucleosynthetic_2002}, a non-rotating Pop~III star with $M = 100~\msun$ is expected to collapse directly into a black hole rather than explode as a PISN. However, we highlight that the primary goal of this work is to investigate the impact of Pop~III supernova on the build-up of an X-ray background, rather than the consequence of a particular stellar evolutionary pathway. Future work will explore stellar masses, explosion energies, and remnant population that are more closely tied to stellar evolution models.

After the explosion, the Pop~III particle is replaced by a supernova remnant particle that persists for $10^4$~yr and emits X-ray photons (Group~6, Table~\ref{tab:bin}), contributing to the local X-ray background. The X-ray energy released is $E_{HN,X} = 100 \times E_{SN,X} = 6 \times 10^{50}$~erg, where $E_{SN,X} = 6 \times 10^{48}~{\rm erg}$ is the X-ray energy from a standard SN remnant \citep[see][and R16]{lopez_using_2011}.

\subsection{Pop~II Stars}
\label{sec:pop2}

Pop~II stars form when metal-enriched gas cools and collapses on an appropriate timescale. A Pop~II particle is created in a gas cell when:
\begin{enumerate}
    \item the hydrogen number density ($\nh$) exceeds $10^4$~\hcc; and
    \item the metallicity exceeds $10^{-4}~\zsun$.
\end{enumerate}
Unlike Pop~III star formation (Section~\ref{sec:pop3}), Pop~II star formation is evaluated at a fine time step, avoiding the expensive clump-finding procedure. The local Pop~II SFR follows the Schmidt relation \citep{schmidt_rate_1959}, $\dot{\rho}_{*} = \epsilon_{SF} \rho_{gas}/t_{ff}$, where $t_{ff} = \sqrt{3 \pi/32G\rho}$ is the free-fall time and $\epsilon_{SF}=0.1$ is the star formation efficiency. Each Pop~II particle represents a star cluster of mass $M_{pop2} \sim 5 \times 10^4~\msun$, with luminosities computed using the stellar population synthesis model of \citet{bruzual_stellar_2003} and a Chabrier IMF \citep{chabrier_galactic_2003}. After $10$~Myr, $10\%$ of the particle mass undergoes SN explosions, injecting energy and metals into the surrounding medium, with $5\%$ of the ejecta mass in metals \citep{kimm_towards_2015}.

\subsection{Effect of Radiation Background}
\label{sec:background}

EUV and X-ray radiation ionise the gas and enhance \hm formation. In Papers~I and II, we studied individual minihaloes exposed to external X-ray and LW backgrounds of varying intensities to identify the conditions under which Pop~III star formation is suppressed or enhanced.

Here we run cosmological volume simulations to self-consistently estimate the background intensities produced by Pop~III and Pop~II star formation, accounting for both local and global contributions. Multiple Pop~III stars can form and radiate within the same simulation volume, contributing to the local background, while our new analytic method (Section~\ref{sec:onthefly}) computes the global background from distant sources, mitigating the limitations of small box sizes relative to the X-ray and LW photon mean free paths.

The ionisation rate of chemical species $\rm i$ in a gas cell is given by,
\begin{equation}
    \frac{\dd x_{i}}{\dd t} = -(1 + f_{ion,i}) \zeta^{\rm i},
    \label{eq:ion}
\end{equation}
and the heating rate due to species $\rm i$ is,
\begin{equation}
    \frac{\dd e_{i}}{\dd t} = f_{heat,i} \Gamma^{\rm i}.
    \label{eq:heat}
\end{equation}
Here, ${\rm i}$ represents one of \hi, \hei, and \heii. $x_{i}$ is the fractional abundance, and $e$ is the internal energy. The terms $\zeta^{\rm i}$ and $\Gamma^{i}$ denote the photoionisation and photoheating rates, respectively, while $f_{ion,i}$ and $f_{heat,i}$ are the secondary ionisation and heating fraction, accounting for the effects of the photoelectrons. Further details on these secondary processes are given in \citet{shull_x-ray_1985}, \citet{ricotti_fate_2002a}, and Paper~I.

The photoionisation rate is divided into contributions from local and global sources,
\begin{equation}
    \zeta^{\rm i} = \zeta_{loc}^{\rm i} + \zeta_{glo}^{\rm i}.
\end{equation}
The local photoionisation rate, arising from local radiation transported within the simulation, is given by
\begin{equation}
    \zeta_{loc}^{\rm i} = \sum_{j=2}^{6} \sigma_{i}^{\rm j} c_{red} n_{j},
    \label{eq:zeta_loc}
\end{equation}
where ${\rm j}$ denotes an photon group (Table~\ref{tab:bin}). Group ${\rm j}=1$ (LW) is excluded in the above equation because it contributes to only \hm-photodissociation. $\sigma_{i}^{\rm j}$ is the ionisation cross-section of species $\rm i$ for photons in group $\rm j$. $n_{j}$ is the number density of photons in the cell. $c_{red} = f_{red} c$~is the reduced speed of light with $f_{red} = 10^{-3}$.

The global photoionisation rate due to the global radiation background is,
\begin{equation}
    \zeta_{glo}^{\rm i} = 4\pi \int_{\it I_{\rm i}/\hp}^{\infty} \frac{\intj_{\nu}}{h_{P} \nu} \sigma_{\rm i}(\nu) \dd \nu,
    \label{eq:zeta_glo}
\end{equation}
where $h_{P}$ is the Planck constant, $I_{\rm i}$ is the ionisation potential (\eg, $I_{H~I} = 13.6$~eV), and $\sigma_{\rm i}(\nu)$ is the ionisation cross-section of species $\rm i$ at frequency $\nu$ taken from \citet{verner_atomic_1996}. The key difference from the local rate is that the global rate is computed using the full spectrum, whereas the local rate is based on discrete photon groups (Table~\ref{tab:bin}). $\intj_{\nu}$ denotes the specific intensity of the radiation background, with its calculation described in Section~\ref{sec:onthefly}.

Heating rates from local and global radiation are calculated analogously,
\begin{equation}
    \Gamma^{\rm i} = \Gamma_{loc}^{\rm i} + \Gamma_{glo}^{\rm i}.
\end{equation}

The \hm fraction ($x_{H_2}$) is determined by the rates of \hm formation ($R$), collisional dissociation ($C_{coll}$), and photodissociation by the global ($k_{glo}$) and local ($k_{loc}$) LW backgrounds,
\begin{equation}
    \frac{\dd x_{H_2}}{\dd t} = Rx_{HI} - (C_{coll} + k_{glo} + k_{loc}) x_{H_2}.
\end{equation}
We refer the reader to Paper~I for the descriptions of $R$, $C_{coll}$, and $k_{glo}$. In Paper~I, only the global LW background was considered, and its \hm-dissociating effect is represented by $k_{glo}$. A new process included in this work is the \hm-photodissociation by the local LW background (photons in Group~1) transported using the M1-closure method. Their contribution to the photodissociation rate in a cell is given by,
\begin{equation}
    k_{loc} = \sigma_{LW} f_{shd} c_{red} n_{1},
    \label{eq:k_loc}
\end{equation}
where $\sigma_{LW} = 2.47 \times 10^{-18}$~cm$^2$ is the effective \hm photodissociation cross-section \citep{katz_interpreting_2017}, $f_{shd}$ is the self-shielding factor \citep{wolcott-green_h2_2019}, and $n_{1}$ is the density of Group~1 photons in the cell. A more detailed discussion of the self-shielding factor can be found in Section~2 of Paper~III.

\subsection{On-the-fly Calculation of Global Radiation Background}
\label{sec:onthefly}

As discussed in Section~\ref{sec:motivation}, improving estimates of the global X-ray background requires: (1) resolving the first sources of radiation (here, Pop~III stars); (2) capturing the feedback loop between X-rays/LW and Pop~III star formation via an on-the-fly background calculation; and (3) including contributions from distant sources. Points (2) and (3) both depend critically on tracking the Pop~III star formation history (SFH). As R16 emphasised, the X-ray feedback loop reinforces over time, making it essential to capture how Pop~III star formation and the background influence each other in real time. Point (3) requires a large box ($L \gtrsim 10~\mpch$), but simultaneously achieving high resolution and large volume is computationally prohibitive.

To address this, we approximate radiation from distant sources analytically by integrating the SFH along the lightcone, thereby capturing both the feedback loop and long-range contributions without simulating an enormous volume.

The on-the-fly calculation rests on two main assumptions.
\begin{enumerate}
    \item The global radiation background is spatially uniform within the zoom-in region — a good approximation because the mean free paths of LW and X-ray photons far exceed the size of the zoom-in region ($0.6~\mpch$).
    \item The SFH in the target region is representative of the entire Universe, motivating our choice of Volume~2 (whose halo mass function matches the cosmic average) as the representative volume. However, given the small box, the first Pop~III star forms artificially late, causing an unphysical jump in the background intensity. To avoid this, we use the analytic estimate from R16 at early times when fewer than 10 Pop~III stars exist in the box ($z > z_{ana}$). The resulting background SED is then applied to all other volumes.
\end{enumerate}

We now describe how the global radiation background ($\intj_{\nu}$ in equation~(\ref{eq:zeta_glo})) is computed; see also Fig.~\ref{fig:schematic}. Following equation~(2) of \citet{haardt_radiative_2012}, the specific intensity at observed frequency $\nu_{o}$ when the box redshift is $z_{o}$ is
\begin{equation}
    \begin{split}
        \intj_{\nu_{o}}(z_{o}) 
        &= \frac{c}{4\pi} \int_{z_{box}}^{z_{start}} \left| \frac{\dd t}{\dd z} \right| \dd z \frac{(1+z_{o})^3}{(1+z)^3} \epsilon_{\nu}(z) e^{-\tau_{\nu}(z)} \hspace{10pt}\\
        & \hspace{3.5cm} [{\rm erg}~{\rm s}^{-1}~{\rm cm}^{-2}~{\rm Hz}^{-1}~{\rm Sr}^{-1}] \\
    \end{split}
    \label{eq:background}
\end{equation}
where $\nu = \nu_{o} (1+z)/(1+z_{o})$. We integrate over sources between $z_{start} = 40$ and $z_{box}$, where $z_{box}$ corresponds to half the size of the zoom-in region (e.g., $0.6~\mpch$ for Volumes~1--5). Radiation from sources within this scale ($z_{o} \leq z \leq z_{box}$) is excluded because it contributes to the local terms $\zeta_{loc}$ and $\Gamma_{loc}$. The integral is evaluated numerically at each coarse time step, and the resulting background is held constant until the next step.


\begin{figure}
    \centering
	\includegraphics[width=0.48\textwidth]{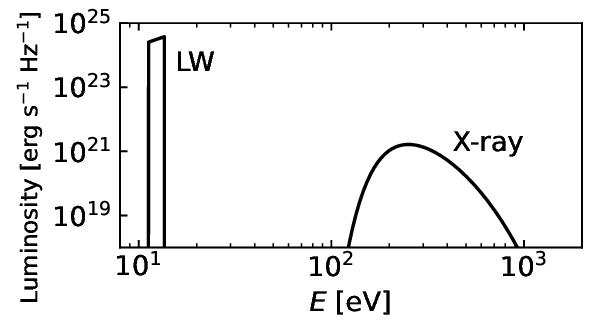}
    \caption{Spectrum of a single Pop~III star.}
    \label{fig:single_spec}
\end{figure}

The optical depth $\tau_{\nu}(z)$, which accounts for absorption of radiation in equation~(\ref{eq:background}), is defined as follows,
\begin{equation}
\tau_{\nu}(z)=
\left\{
	\begin{array}{ll}
		\dd s/\ell_{mean} & {\rm (LW)} \\
		\alpha_{\nu}(z) \dd s & \text{(X-ray)} \\
	\end{array}
\right.,
\label{eq:tau}
\end{equation}
where $\dd s = c \dd t$ is the path length over the subinterval at $z$. For LW radiation, $\ell_{mean}$ is the effective mean free path, taken to be 1/150 of the particle horizon following R16,
\begin{equation}
    \ell_{mean} = \frac{t_{univ}c}{150},
\end{equation}
where $t_{univ}$ is the age of the Universe at $z$. For X-ray radiation, the opacity is
\begin{equation}
    \begin{split}
        \alpha_{\nu}(z) &= n_{\hi}(z) \sigma_{\hi}(\nu) +n_{\hei}(z) \sigma_{\hei}(\nu) \\
        &= n_{H}(z) x_{\hi}(z) \sigma_{\hi}(\nu) + n_{He}(z) x_{\hei}(z) \sigma_{\hei} (\nu) \hspace{10pt} [{\rm cm}^{-1}],
    \end{split}
    \label{eq:opacity}
\end{equation}
where $n_{i}(z)$ represents the number density of species $\rm i$. Here, we neglect the contribution from \heii. The hydrogen density at $z$ is calculated as,
\begin{equation}
    n_{H}(z) = h^2 (1+z)^3 \rho_{crit} \Omega_{b} \frac{X}{m_{H}},
    \label{eq:nh}
\end{equation}
where $\rho_{crit} = 1.88 \times 10^{-29}~{\rm g}~{\rm cm}^{-3}$ is the critical density of the Universe, $X=0.76$ is the hydrogen mass fraction, and $m_{H}$ is the mass of a hydrogen atom. The helium number density is,
\begin{equation}
    n_{He}(z) = n_{H}(z)\frac{Y}{X}\frac{1}{4},
    \label{eq:nhe}
\end{equation}
where $Y=0.24$ is the helium mass fraction. The neutral fractions $x_{\hi}(z)$ and $x_{\hei}(z)$ are approximated by their values at $z_{o}$ for simplicity. At $z = 9$, the neutral fraction is $\sim 0.8$--$0.9$. This approximation may underestimate the optical depth at high redshifts by $\sim 20\%$, leading to a corresponding overestimate of the X-ray background intensity. 


\begin{figure*}
    \centering
	\includegraphics[width=0.95\textwidth]{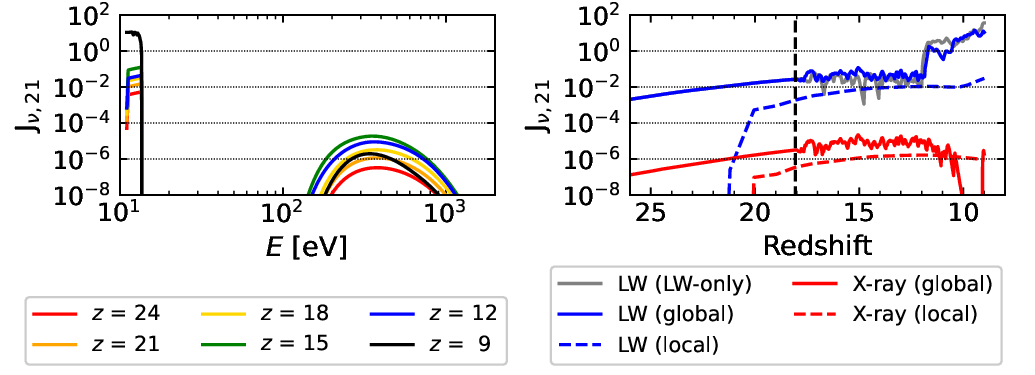}
    \caption{Radiation background intensities in units $\intj_{21}$ estimated from the V2X simulation. The left panel shows the spectra of the global radiation backgrounds at several selected redshifts. The LW intensity ranges from $\sim 10^{-2}$ to $\sim 10^{-1}$, while the peak X-ray intensity reaches $\sim 10^{-5}$ at $z = 15$. The right panel presents the evolution of X-ray at $E = 360$~eV (red) and LW (blue) intensities with redshift, with global backgrounds in solid lines and local backgrounds in dashed lines. The local backgrounds are estimated from the photon number density following equation~(18) of \citet{trenti_formation_2009}. The vertical dashed line marks $z_{ana}$ for this simulation (see Section~\ref{sec:onthefly} and equation~(\ref{eq:npop3}). For comparison, the grey line shows the LW intensity in the LW-only simulation (V2LW). With the onset of Pop~II star formation at $z \sim 12$, the LW intensity rises rapidly. In contrast, the decline in X-ray intensity is more gradual around the same redshift, as ongoing Pop~III star formation is sustained by positive X-ray feedback.
    Near the end of the simulation ($z \sim 9$), a single Pop~III star forms in a pristine region, briefly increasing the X-ray background again.}
    \label{fig:global_spec}
\end{figure*}

The emissivity (equation~(\ref{eq:background})) at frequency $\nu$ is given by
\begin{equation}
    \begin{split}
        \epsilon_{\nu}(z) &= j_{\text{pop3}}(\nu) n_{pop3}(z) + j_{\text{pop2}}(\nu) \npoptwo(z) \\
        &\hspace{4cm} [{\rm erg}~{\rm s}^{-1}~{\rm cm}^{-3}~{\rm Hz}^{-1}],
    \end{split}
    \label{eq:emissivity}
\end{equation}
where $j(\nu)$ is the specific luminosity (${\rm erg}~{\rm s}^{-1}~{\rm Hz}^{-1}$) of a single star particle (Pop~III or Pop~II), and $n(z)$ is the number density of stars at $z$. The Pop~III spectrum $j_{\text{pop3}}(\nu)$ has two components: LW and X-ray. The X-ray spectrum follows the thermal bremsstrahlung formula \citep{rybicki_radiative_1986}
\begin{equation}
    \begin{split}
        \epsilon_{\nu}^{\rm ff} &= 6.8 \times 10^{-38} Z_{ch} n_{e} n_{i} T^{-1/2} e^{-h_{P}\nu / k_{B}T} \bar{g_{ff}} e^{-\tau_{SN}} \\
        &\hspace{4.5cm} [{\rm erg}~{\rm s}^{-1}~{\rm cm}^{-3}~{\rm Hz}^{-1}],
    \end{split}
    \label{eq:brem}
\end{equation}
where $g_{ff} = \left( \frac{3}{\pi} \frac{k_{B}T}{h_{P} \nu} \right)^{1/2}$ is the averaged Gaunt factor \citep[see][]{rybicki_radiative_1986}. $k_{B}$ is the Boltzmann constant, and the temperature of the supernova remnant is $T=10^6$~K. The term $e^{-\tau_{SN}}$ accounts for absorption in the remnant, with $\tau_{SN}$ defined as $N_{col,H} \sigma_{H I}(\nu)$, where $N_{col,H} = 10^{21}~{\rm cm}^{-2}$ is the hydrogen column density of the supernova remnant\footnote{We also use $N$ for the number of Pop~III stars in equation~(\ref{eq:npop3}). To indicate column density, we attach `col' in the subscript.}, as inferred from the observations of \citet{lopez_using_2011}. We assume $Z_{ch}=1, n_{e}=1$, and $n_{i}=1$. The spectrum is normalised such that
\begin{equation}
    \begin{split}
        \int_{0.2~{\rm keV}/h_{P}}^{2.0~{\rm keV}/h_{P}} \epsilon_{\nu}^{\rm ff} \dd \nu
        &= \frac{E_{HN,X}}{t_{pop3}}\\
        &= \frac{\alpha_{HN} E_{SN,X}}{t_{pop3}},
    \end{split}
\end{equation}
where $\alpha_{HN}$ is the boost factor of the HN X-ray energy. 

As a side note, the opacity (equation~(\ref{eq:opacity})) is recalculated whenever the global background is updated, rather than being stored in memory and incrementally evolved. The cosmic hydrogen and helium densities are estimated directly from cosmology (equations~(\ref{eq:nh}) and (\ref{eq:nhe})). Although the neutral fractions, $x_{\hi}(z)$ and $x_{\hei}(z)$, are time-dependent quantities determined by ionising radiation, we approximate them using the values at the current redshift, $x_{\hi}(z_{o})$ and $x_{\hei}(z_{o})$. Since these quantities can be evaluated directly at each update, no additional memory is required to store their past evolution. The emissivity, on the other hand, depends on the number densities of Pop~III and Pop~II stars, $n_{pop3}(z)$ and $n_{pop2}(z)$ (equation~(\ref{eq:emissivity})), which constitute the key time-dependent quantities in our model. We therefore store their evolution in memory. When new stars form, the corresponding information is appended to a time-ordered table, which is subsequently used to recalculate the global radiation background.

As outlined in Section~\ref{sec:pop3}, a Pop~III star particle enters the supernova remnant phase after its main-sequence lifetime and emits 100 times more total and X-ray energies than a normal SNe ($\alpha_{HN} = 100$) for $10^4$~yr. X-ray photons emitted in this way contribute to the local X-ray background. When estimating the global X-ray background, however, we adopt a larger value of $\alpha_{HN} = 1000$ to account for Pop~III multiplicity. The smaller boost factor used for the local X-ray sources is motivated by computational efficiency. As shown in the right panel of Fig.~\ref{fig:global_spec}, the local X-ray background is weaker than the global one\footnote{Since photons propagate with a reduced speed of light, $c_{red} = 10^{-3}c$, they remain in the target region for longer and the local radiation intensity is overestimated. Adopting $\alpha_{HN} = 1000$ would increase the local X-ray intensity which is comparable to the global X-ray intensity. However, the corresponding photoionisation and photodissociation rates are reduced by the same factor \citep[see also equations~(\ref{eq:zeta_loc}) and (\ref{eq:k_loc})]{rosdahl_ramses-rt_2013}. Consequently, the impact of the local X-ray background is smaller than suggested by Fig.~\ref{fig:global_spec}.}. Increasing $\alpha_{HN}$ for the local sources therefore has little impact on the results, but would inject more energy into a small number of cells over a short timescale ($10^4$~yr), reducing the simulation time step and increasing the compuation cost.

The LW spectrum is taken from \citet{schaerer_properties_2002} and illustrated in Fig.~\ref{fig:single_spec}. Note that the duration of X-ray emission during the supernova phase ($\sim 10^4~{\rm yr}$) is much shorter than the main-sequence lifetime ($2.77~\myr$). For the global background, however, we consider the time-averaged effect over all Pop~III stars; accordingly, we distribute the X-ray emission uniformly over the main-sequence lifetime rather than concentrating it at the moment of explosion. The number density of Pop~III stars\footnote{Strictly speaking, this is the number density of minihaloes that can host Pop~III stars. Since we assume one Pop~III star per halo, these values are equivalent.} is
\begin{equation}
    n_{pop3}(z) = 
    \left\{
	\begin{array}{ll}
		N_{pop3}(z)/L_{target}^3 & (z < z_{ana}) \\
		f_{duty} \int_{{\it M}_{crit,1}}^{\infty} \left( \frac{\dd n}{\dd \ln{M}} \right) \frac{1}{M}\dd M &  (z > z_{ana})
	\end{array}
\right..
\label{eq:npop3}
\end{equation}
$N_{pop3}(z)$ is the number of Pop~III stars in the target region. At high redshifts, when few or no Pop~III stars are present, the emissivity from the first line is highly uncertain and episodic. We therefore use the analytic estimate in the second line for $z > z_{ana}$, integrating the halo mass function \citep{diemer_colossus_2018} above the critical mass $M_{crit,1} = 2 \times 10^6~\msun$ and multiplying by the duty cycle $f_{duty} = t_{pop3}/t_{univ}$, following R16. Once the number of Pop~III stars in the simulation exceeds 10 ($z < z_{ana}$), $n_{pop3}(z)$ is simply $N_{pop3}(z)$ divided by the target volume.

The contribution of Pop~II stars to the emissivity is treated similarly, with two key differences. First, we do not estimate $\npoptwo$ analytically at high redshift, so the LW background exhibits a jump at the formation of the first Pop~II star in the box — an artefact, since the global background should evolve smoothly. This leads to a sharper transition from Pop~III to Pop~II star formation than is physical, and we plan to improve this treatment in future work. Second, we include only Pop~II LW radiation, neglecting their contribution to the ionising UV and X-ray backgrounds. This simplification is justified by our focus on the pre-reionisation era \citep[$z_{re} = 7.68$,][]{planck_collaboration_planck_2020}, during which ionising UV has not yet permeated the IGM, and by our goal of evaluating the results of R16 and Paper~I with the X-rays from Pop~III supernovae only. The role of Pop~II supernovae and X-ray binaries will be explored in future work.

Finally, $j_{pop2}(\nu)$ is computed for each Pop~II particle using its age, mass, and metallicity, based on the population synthesis model of \citet{bruzual_stellar_2003}. The resulting global X-ray and LW backgrounds from the on-the-fly method are shown in Fig.~\ref{fig:global_spec} and described in the following section.


\begin{figure*}
    \centering
	\includegraphics[width=0.9\textwidth]{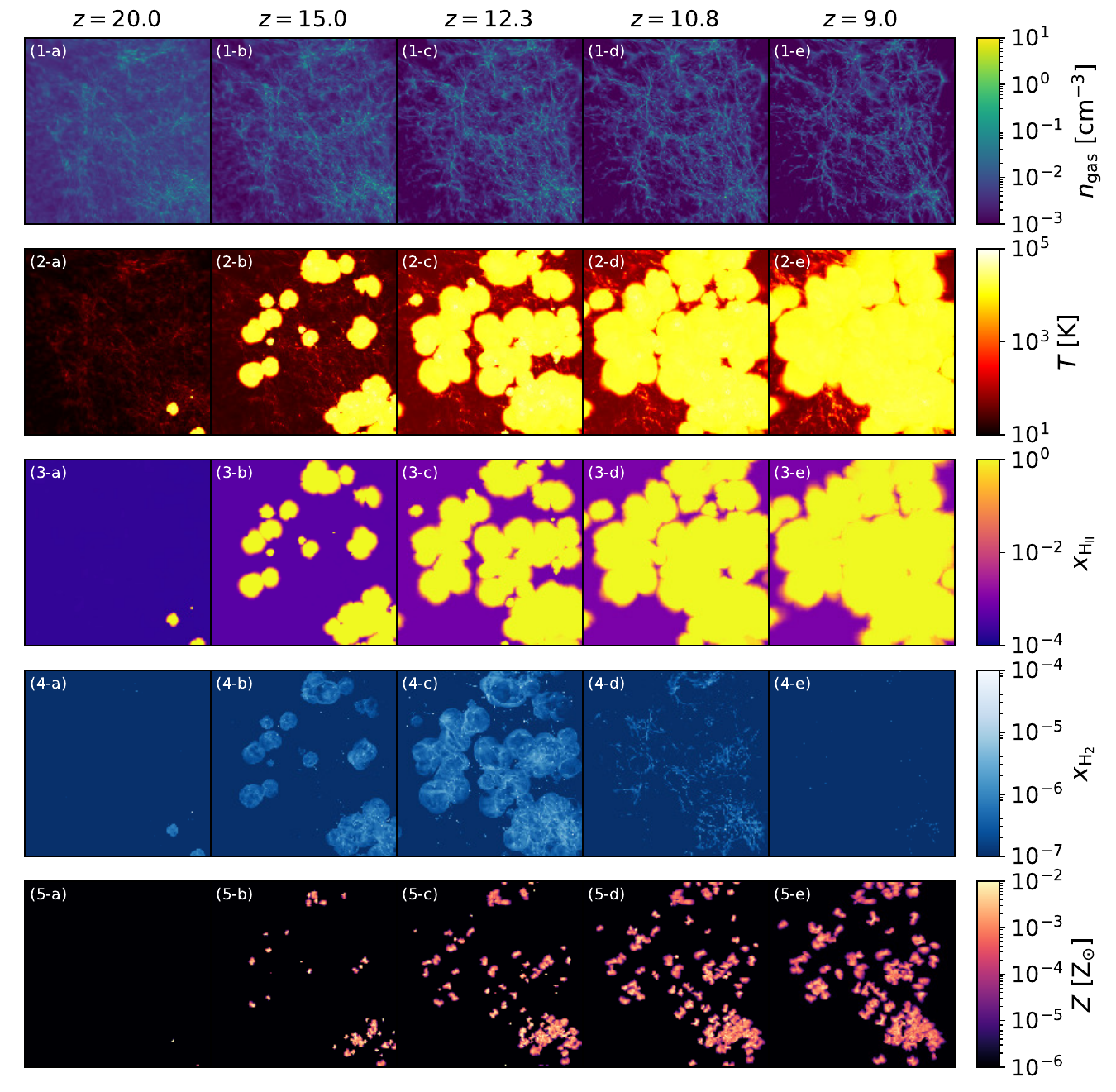}
    \caption{Evolution of gas in the target volume of V2X at $z = 20.0, 15.0, 12.3, 10.8$, and $9.0$ (from left to right, labelled a to e). The rows show (from top to bottom, labelled 1 to 5) the gas density, temperature, \hii fraction, \hm fraction, and metallicity. As shown in the second and third rows, \hii bubbles (yellow) form around Pop~III stars, and the volume covered by these regions increases with time as the \hii bubbles grow and merge. In contrast, the IGM is mildly heated and ionised by the global X-ray background ($T \sim 100$~K and $x_{e} \sim 5 \times 10^{-3}$). The fourth row shows the formation of \hm shells around \hii bubbles. However, an increase in the IGM \hm fraction due to X-rays is not evident, as the LW background dominates. As the LW intensity increases significantly below $z \sim 12$, the \hm shells are destroyed and the \hm fraction remains extremely low across the volume (in Panel 4-e, $x_{\text{\hm}} < 10^{-7}$). As stars explode, the volume polluted with metals increases with time (bottom row). }
    \label{fig:overview}
\end{figure*}


\begin{figure*}
    \centering
	\includegraphics[width=0.9\textwidth]{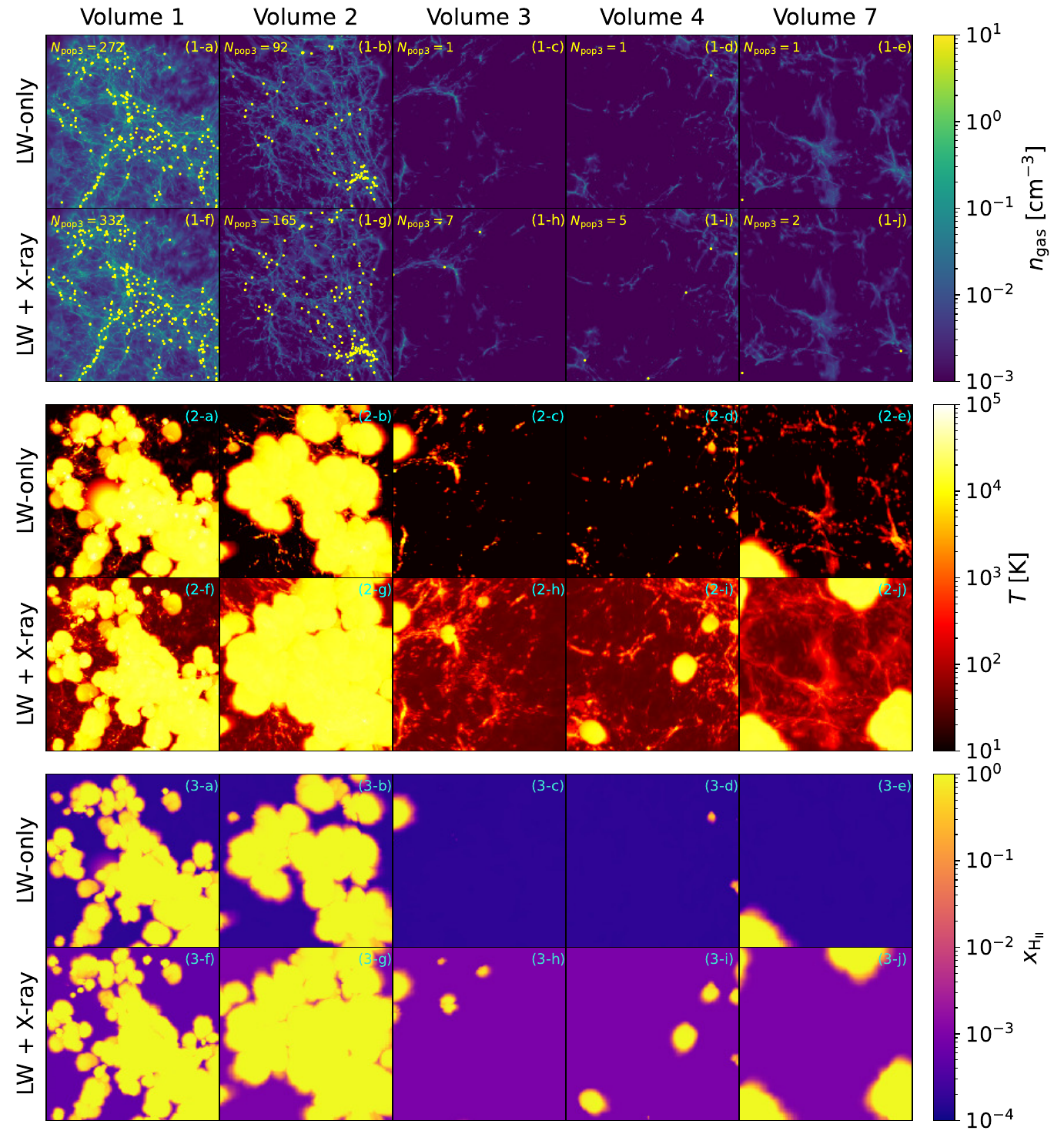}
    \caption{Gas density, temperature, and \hii fraction (top to bottom) of the target volumes. From left to right, different volumes are shown. For Volume~1, results are presented at $z \sim 13.56$ and $13.94$; for the other volumes, results are shown at $z = 9$. From Panels 1-a to 1-j, Pop~III stars and their remnants are shown in yellow dots. In addition, at the top left corner of each of the ten panels, we present the number of Pop~III stars that have formed in the target region. In Volume~3, fewer Pop~III stars are shown than the given value of $N_{pop3}$ as some stars drift and escape the target region. For each property, the LW-only case (top row) is compared with the X-ray cases (bottom row). Each panel displays the maximum value along the line of sight. While the gas density maps in the LW-only and X-ray simulations are indistinguishable, the temperature and ionisation maps show clear differences. Enhanced Pop~III star formation in the X-ray runs (see each bottom row) produces hot \hii regions that cover a bigger volume (yellow regions). In addition, the global X-ray background mildly heats and ionises the IGM, raising the background temperature to $T$ ($\sim 100$~K) and the \hii fraction to $x_{H~II}$ ($\sim 10^{-3}$).}
    \label{fig:map1}
\end{figure*}

\begin{figure*}
    \centering
	\includegraphics[width=0.95\textwidth]{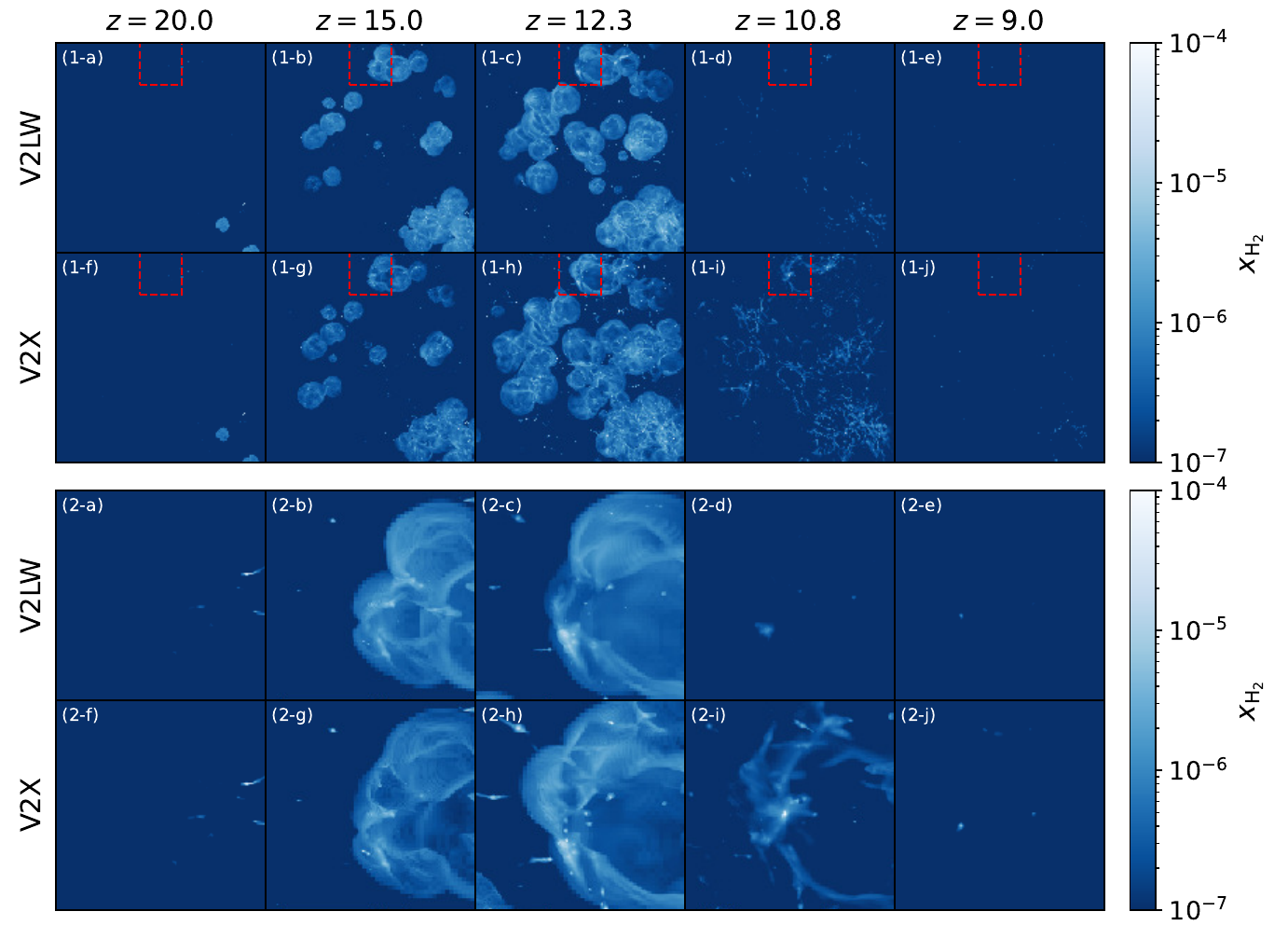}
    \caption{\hm fraction of the target region of Volume~2 (top grid) and its sub-volume (bottom grid). From left to right, we show snapshots at $z = 20.0, 15.0, 12.3, 10.8$, and $9.0$. In each grid, the LW-only case (V2LW, top row) is compared with the X-ray case (V2X, bottom row). The red squares in the top grid mark the sub-volumes shown in the bottom grid. In close-up views, the more numerous dots in the X-ray case (e.g., Panels~1-h, 1-i, 2-h, and 2-i) indicate higher \hm fractions in DM haloes, although the fractions remain low in the IGM. Panels 1-d and 1-i highlight the large-scale X-ray effect: in V2X, the \hm fraction remains higher in the DM haloes and filaments despite the strong LW background.}
    \label{fig:map_xhm}
\end{figure*}


\section{Results}
\label{sec:xray}

\subsection{Overview}

Fig.~\ref{fig:global_spec} shows the global radiation background computed on-the-fly. Both the LW and X-ray intensities rise until $z \sim 15$ and remain nearly constant until $z \sim 12$. With the onset of Pop~II star formation, the LW intensity rises rapidly to $\sim 10^1~\intj_{21}$, a strong-LW regime (Paper~I) that suppresses Pop~III star formation via \hm photodissociation. The X-ray intensity, by contrast, decreases only gradually, as positive X-ray feedback sustains some Pop~III star formation against the intensifying LW background. As LW radiation strengthens further and gas becomes metal-enriched, star formation transitions from Pop~III to Pop~II, causing a sharp drop in X-ray intensity at $z \sim 10$ (since only Pop~III stars emit X-rays in our model). As noted in Section~\ref{sec:onthefly}, the rapid rise of the LW background at $z \sim 12$ is artificially steep because it corresponds to the formation of the first Pop~II star in the small box, and we plan to improve this treatment in future work. In this context, \citet{incatasciato_modelling_2023} provides a useful point of comparison. Their LW background, derived from both Pop~III and Pop~II stars in a larger cosmological volume (22.72~Mpc/$h$), reaches intensities similar to those found in our work. For example, the LW intensity is J$_{21}\sim 10^{-1}$ at $z \sim 15$ and J$_{21}\sim 10$ at $z \sim 9$. However, their LW background evolves more smoothly with redshift, in contrast to the steep increase seen in our model at $z \sim 12$.

The local LW background (blue dashed line) does not exhibit a sharp increase in its intensity at $z \sim 12$ seen in the global LW background. We speculate that this is a consequence of the reduced speed of light approximation ($f_{red} = 10^{-3}$), which prevents the LW photons emitted by Pop~III stars from escaping the target region efficiently and therefore leads to an overestimate of the local LW intensity. Around $z \sim 12$, Pop~II stars produce only a comparable amount of LW radiation (even if they are greater in mass, e.g., $M_{gal} \sim 10^6~\msun$) to the accumulated Pop~III contribution, and therefore do not induces a sharp increase in the local background. A more noticeable rise appears at $z \sim 10$, when the total Pop~II mass exceeds $10^7~\msun$.

Fig.~\ref{fig:overview} shows the gas evolution in Volume~2 (V2X), a mean-density region. The first row shows the growth of filaments and DM haloes. The second and third rows illustrate two modes of radiative feedback: local stellar feedback, which heats and ionises gas to form \hii regions, and the global X-ray background (Section~\ref{sec:onthefly}), which mildly heats and ionises the nearly neutral gas between bubbles, reaching $T \sim 100$~K and $x_{\hii} \sim 5 \times 10^{-3}$ by $z = 9$ (Panels 2-e and 3-e). The fifth row tracks the growth of the metal-enriched volume. The fourth row shows the formation of \hm shells around Pop~III stars \citep{ricotti_fate_2002a,ricotti_fate_2002b} that overlap and increase the \hm fraction until $z \sim 12$ (panels 4-a to 4-c), after which the rapidly rising LW intensity destroys these shells and suppresses \hm to low values (4-d and 4-e). The remainder of this section compares simulations with and without X-rays to isolate the role of X-ray heating and ionisation.

\subsection{Global Radiation Background Effects on Primordial Gas}
\label{sec:global_feedback}

Fig.~\ref{fig:map1} compares gas properties in simulations without (top row in each grid) and with (bottom row in each grid) X-ray radiation, highlighting X-ray effects. Each column shows a different sub-volume (overdense to underdense from left to right). The temperature and \hii fraction maps (Grids~2 and 3) show that X-rays increase the overall ionization and temperature of the IGM. X-rays increase the \hii regions volume filling factor because of the increased number of Pop~III stars (i.e., local UV feedback) and the IGMs between bubbles is mildly heated and ionised. This enhancement is evident in Volume~2 (mean density) and in sparsely populated (underdense) regions such as Volumes~3 and 4.

In contrast, Volumes~1 and 7 show little difference in the spatial extent of the \hii regions between the LW-only and X-ray simulations. The origin of this environmental effect is discussed in detail in the companion paper PR26b (Section~3.2). In brief, the key factor is the relative intensity of X-ray vs LW backgrounds. When Pop~III-forming minihaloes ($M \gtrsim 10^6~\msun$) virialise at intermediate redshift between $z \sim 16$ and $z \sim 12$, the X-ray background is near its maximum intensity, while the LW background remains moderate. However, in Volume~1 (overdense region), such haloes form early ($z \gtrsim 16$), when the X-ray background is still weak. In Volume~7 (underdense region), on the other hand, they appear later ($z \sim 12$), when the strong LW background dominates. As a result, the X-ray effects are minimal in both cases. {\it We conclude that the positive effect of X-rays on Pop~III formation is highest in mean density to moderately underdense regions of the IGM}.

Unlike X-ray heating and ionisation, enhanced \hm formation in gas phase (equation~(\ref{eq:hminus})) is not evident in Row~4 of Fig.~\ref{fig:overview}. Although the IGM temperature and \hii fraction increases with time ($T \sim 100$~K and $x_{\text{\hii}} \sim 5 \times 10^{-3}$; Rows~2 and 3), the \hm fraction of the IGM remains low. This is because the LW background is several orders of magnitude stronger than the X-ray background (Fig.~\ref{fig:global_spec}) and efficiently photodissociates \hm in the diffuse gas. However the formation rate of H$_2$ scales with the density squared while the dissociation is proportional to the density. Hence, for the dense gas within DM haloes, \hm formation can outpace negative LW feedback, enabling Pop~III star formation. Consequently, the global \hm fraction in the IGM does not effectively trace positive X-ray feedback on Pop~III star formation.

To examine the \hm enhancement by X-rays, we compare V2LW and V2X in Fig.~\ref{fig:map_xhm}. At early times ($z \gtrsim 15$), the X-ray background is weak and the differences between the two simulations are minimal. The contrast becomes more apparent at $z = 12.3$ (Panels~c and h) and $10.8$ (Panels~d and i), especially in Grid~2, where bright clumps are clearly visible in the X-ray simulation (see 2-h and 2-i), indicating higher \hm fractions within DM minihaloes. Despite the \hm enhancement in haloes, the IGM \hm fraction in both runs is similar with values reaching $10^{-4}$ in relic \hii regions and in shells around ionization fronts \citep{ricotti_feedback_2001}. On larger scales (Grid~1), X-ray effects are subtle, though at $z = 10.8$ a modest increase appears in dense filaments. At this epoch, the X-ray feedback loop sustains Pop~III star formation, allowing weak X-rays to partially offset LW feedback. By $z = 9$, Pop~III star formation is nearly quenched, the X-ray intensity decreases, and the \hm fraction drops below $10^{-7}$ throughout the volume.


\begin{figure}
    \centering
	\includegraphics[width=0.48\textwidth]{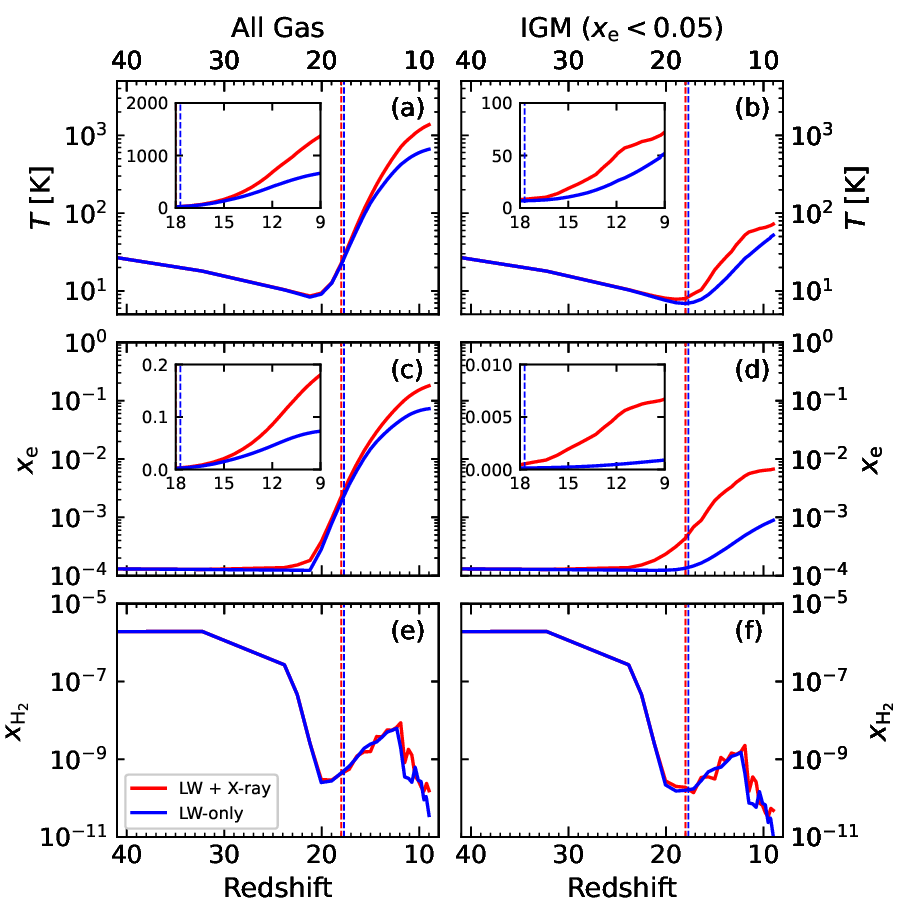}
    \caption{Volume-weighted mean temperature, electron fraction, and \hm fraction (from top to bottom) of V2. In the left column, we average all gas cells in the target volume. In the right column, on the other hand, only gas cells with $x_{e} < 5 \times 10^{-2}$ (i.e., those ionised by the weak global X-ray) are considered to separate the IGM from \hii bubbles, thereby isolating the effects of the global background on the IGM. The primary results are robust to the choice of threshold (0.01 and 0.1). In each panel, the results of V2X are plotted in red, and the result of the LW-only counterpart in blue. Sub-panels in the top and middle rows display the same data on a linear scale to highlight differences between the simulations. Although the differences in the temperature and \hm fraction are within a factor of $\sim 2$, the IGM electron fraction (middle right panel) exhibits a larger difference (by a factor of $\sim 10$).}
    \label{fig:global}
\end{figure}


\begin{figure}
    \centering
	\includegraphics[width=0.48\textwidth]{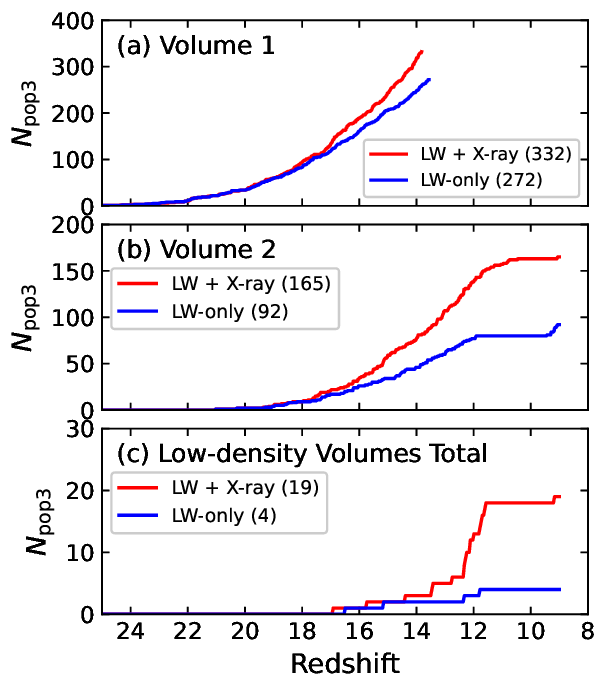}
    \caption{Numbers of Pop~III stars in the X-ray (red) and LW-only (blue) simulations as a function of redshift. In each panel, we present the number of stars at the end of the simulations. Panels~a and b present the results for Volume~1 and Volume~2, respectively. Panel~c shows the total number of Pop~III stars in all low-density volume simulations (from Volume~3 to Volume~10).}
    \label{fig:npop3}
\end{figure}

Fig.~\ref{fig:global} examines these effects quantitatively for Volume~2, showing volume-weighted quantities as a function of redshift. Panel~a shows that the mean gas temperature is consistently higher in the X-ray run, reflecting the combined contributions of stellar feedback and the global X-ray background. At $z = 9$, mean temperatures in the target region are $\sim 700$~K (blue) and $\sim 1400$~K (red) without and with X-rays, a factor-of-two increase driven largely by enhanced Pop~III star formation — highlighting the substantial role of local UV stellar feedback, which was not included in R16 and Paper~I.

Panel~b restricts the same comparison to gas outside \hii regions ($x_{e} < 0.05$). The X-ray background heats the IGM only mildly: IGM temperatures at $z = 9$ are $\sim 50$~K and $\sim 70$~K in V2LW and V2X, respectively. This modest contrast ($70/50 = 1.4$) arises because most photoelectron energy goes into ionisation rather than heating in a neutral medium \citep{shull_x-ray_1985, ricotti_fate_2002a}, with $f_{heat} \sim 0.1$ when $x_{e} \sim 10^{-4}$ (equation~(\ref{eq:heat})). Despite X-ray ionisation, the IGM remains nearly neutral ($x_{e} \lesssim 10^{-2}$), limiting heating efficiency. A stronger X-ray background that substantially ionises the IGM would yield a larger temperature contrast as $f_{heat}$ approaches unity.

The mean electron fraction (Panels~c and d) reflects the secondary ionisation process. Including all gas (Panel~c), the fraction rises by a moderate factor of $\sim 2$ due to \hii bubbles. Restricting to the nearly neutral IGM (Panel~d), it rises by a factor of $\sim 8$ ($7 \times 10^{-3}$ versus $9 \times 10^{-4}$), consistent with the high secondary ionisation efficiency in a neutral medium ($f_{ion} \gtrsim 10$, Paper~I).

We find some differences between our results and those of \citet{machacek_effects_2003}. In this work, the X-ray-irradiated IGM reaches a maximum temperature of $T \sim 70$~K and an electron fraction of $x_{e} \sim 7 \times 10^{-3}$. In contrast, the strongest X-ray background considered by \citet{machacek_effects_2003} produces $T \sim 10^3$~K and $x_{e} \sim 10^{-2}$. Although both studies obtain a comparable degree of ionisation, the IGM temperature in \citet{machacek_effects_2003} is an order of magnitude higher, implying more efficient X-ray heating. One possible explanation is the difference in the adopted X-ray spectra. Their spectrum peaks at $E \sim 10^3$~eV and extends to higher energies ($E = 7.6$~keV) owing to the shallow power-law slope ($\alpha=1$), whereas our spectrum is substantially softer (Fig.~\ref{fig:global_spec}). These spectral differences may contribute to the differing thermal evolution of the IGM, and consequently, to the differing strength of X-ray feedback on Pop~III star formation.

Despite the clear differences in temperature and electron fraction, the \hm fraction varies little between the two runs (Panels~e and f), consistent with Fig.~\ref{fig:map_xhm}. LW radiation efficiently photodissociates \hm in low-density gas, while \hm is enhanced in filaments and haloes. Since low-density gas dominates the volume, the volume-weighted \hm fractions differ only marginally.

Nonetheless, higher \hm fractions within DM haloes in the X-ray run lower the critical mass, enabling more haloes to host Pop~III stars. This positive feedback is illustrated in Fig.~\ref{fig:npop3}, which shows the X-ray-driven increase in Pop~III star counts for Volumes~1, 2, and the combined low-density volumes (3--10). The fractional enhancement is weakest in the overdense region (Volume~1) and strongest in underdense environments (Panel~c). A detailed analysis of X-ray effects on the critical mass and Pop~III number density is presented in the companion paper PR26b.

In summary, the X-ray background mildly heats and ionises the IGM ($T_{IGM} \sim 10^3$~K, $x_{e} \sim 7 \times 10^{-3}$). Its impact on the volume-weighted \hm fraction is subtle, but through its density-dependent enhancement of \hm formation, it partially offsets LW negative feedback in dense gas at $z \lesssim 12$, sustaining higher \hm fractions in filaments and DM haloes and thereby boosting Pop~III star formation — with the magnitude varying across environments.


\begin{figure}
    \centering
	\includegraphics[width=0.48\textwidth]{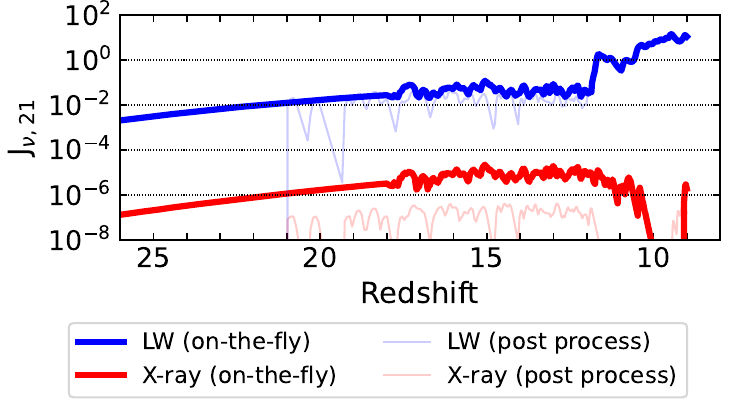}
   	\includegraphics[width=0.48\textwidth]{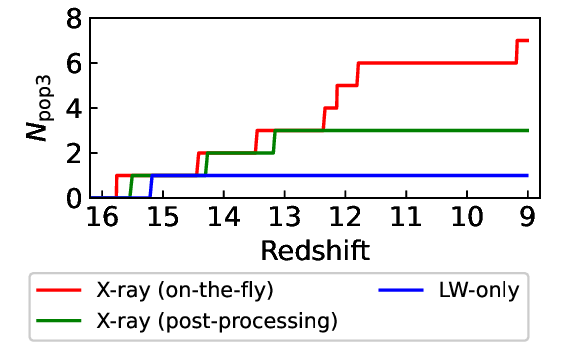}

    \caption{\textbf{Top:} LW (blue) and X-ray (red) radiation backgrounds from the on-the-fly (thick solid lines) and post-processing (thin solid lines) approaches. The format follows the left panel of Fig.~\ref{fig:global_spec}. The post-processed X-ray background is weaker by 1--2 orders of magnitude. \textbf{Bottom:} Number of Pop~III stars in Volume~3 without X-rays (blue, V3LW), with the post-processed X-ray background (green), and with the on-the-fly X-ray background (red).}

    \label{fig:post}
\end{figure}

\subsection{Comparing On-the-fly and Post-processing Methods}
\label{sec:compare}

As established in Section~\ref{sec:motivation}, Pop~III stars and the X-ray background interact in a positive feedback loop (when X-ray heating is sub-dominant): increased Pop~III star formation raises the X-ray background, which in turn promotes further Pop~III formation (R16). We incorporate this loop by computing the X-ray background on-the-fly (Section~\ref{sec:onthefly}). Here we quantify how this approach compares to post-processing, in which X-ray effects are applied a posteriori using the SFH from a simulation that does not include them \citep{xu_heating_2014, ahn_spatially_2015, xu_x-ray_2016}. Specifically, we derive the post-processed X-ray background from the Pop~III SFH ($\npop(z)$; equations~(\ref{eq:emissivity}) and (\ref{eq:npop3})) of the LW-only run (V2LW), and apply it to Volume~3 — mimicking the approach in which an LW background is included self-consistently while X-ray effects are added in post-processing.

The top panel of Fig.~\ref{fig:post} shows that the post-processed X-ray background is weaker by $\sim 1$--$2$ orders of magnitude than the on-the-fly estimate, because it misses the reinforcing feedback loop. As a result, it underestimates the number of Pop~III-forming minihaloes. Applied to Volume~3 — where the X-ray enhancement is most pronounced — the post-processed background yields only three Pop~III stars, compared to seven in the on-the-fly case (bottom panel). We therefore conclude that neglecting the X-ray feedback loop substantially suppresses Pop~III star formation, and we expect this finding to hold generally across environments.


\section{Discussion and Summary}
\label{sec:discussion}

\subsection{Impact on Reionisation and the Optical Depth of the IGM}
\label{sec:tau}

\begin{figure}
    \centering
	\includegraphics[width=0.48\textwidth]{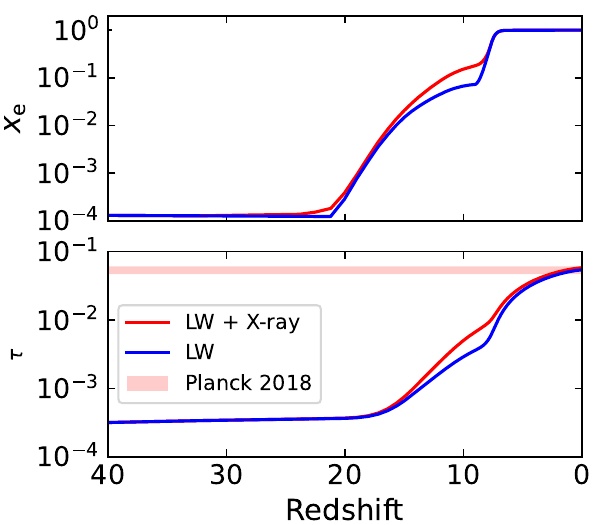}
    \caption{\textbf{Top:} Volume-weighted mean electron fractions of V2X (red) and V2LW (blue). The electron fraction remains elevated at $z \sim 8$--$20$ in the X-ray run. \textbf{Bottom:} Optical depth to Thomson scattering (Eq.~\ref{eq:opt_depth}). The X-ray run yields a higher $\tau$ throughout, though the difference is within a factor of 2. The two curves converge toward $z = 0$ as the pre-$z \sim 9$ contribution is small relative to the final value of $\tau$. In both cases, the total optical depths (0.058 and 0.054) are within the Planck uncertainty \citep[shaded region,][]{planck_collaboration_planck_2020}.}
    \label{fig:tau}
\end{figure}

An X-ray background partially ionises the IGM before reionisation \citep{haiman_radiative_2000, ricotti_x-ray_2004, ricotti_x-ray_2005}, thereby raising the Thomson scattering optical depth. To quantify this effect, we estimate $\tau_e$ for the two Volume~2 simulations. We first compute the volume-weighted mean electron fraction as a function of redshift. Since the simulations end at $z = 9$, we extrapolate the ionisation history using the standard tanh reionisation model \citep{lewis_cosmological_2008,planck_collaboration_planck_2020},
\begin{equation}
    x_{e} = \frac{1}{2} \left[ 1 + \tanh{ \left( \frac{y(z_{re})-y(z)}{\Delta y} \right) } \right],
\end{equation}
where $z_{re} = 7.68$, $y(z) = (1+z)^{3/2}$, and $\Delta y = \frac{3}{4} (1+z_{re})^{1/2}$. The top panel of Fig.~\ref{fig:tau} confirms that the electron fraction is elevated throughout $z \sim 8$--$20$ in the X-ray run.


The bottom panel of Fig.~\ref{fig:tau} shows the optical depth \citep{planck_collaboration_planck_2020},
\begin{equation}
    \tau_{e} (z) = n_{H}(0) c \sigma_{T} \int_{z_{max}}^{z} | \dd z |  x_{e}(z) \frac{(1+z)^2}{\sqrt{\Omega_{m}(1+z)^3 + \Omega_{\Lambda}}},
    \label{eq:opt_depth}
\end{equation}
where $n_{H}(0)$ is the mean hydrogen density at $z = 0$. Rather than integrating from $z = 0$ as in most studies, we integrate from $z_{max} = 120$ (the simulation start) to highlight the time-dependent contribution of X-rays. The two cases diverge at $z \approx 18$, coinciding with the onset of Pop~III star formation, after which $\tau_{e}$ remains higher in the X-ray run. Between $z_{max} = 120$ and $z_{re} = 7.68$, the optical depths are $0.010$ (X-ray) and $0.006$ (LW-only), an increase of $\delta\tau_e \approx 0.004$ ($\sim$150\%) due to X-ray pre-reionisation. Below $z_{re}$, rapid full reionisation contributes similarly in both cases, yielding final total optical depths of $\tau_{e} \approx 0.058$ (X-ray) and $\tau_{e} \approx 0.054$ (LW-only) — both consistent with the Planck measurement $\tau_{e} = 0.054 \pm 0.007$ \citep{planck_collaboration_planck_2020}.

X-rays from Pop~III supernovae thus make only a minor contribution to the total Thomson optical depth. Nevertheless, between $z \sim 10$ and $20$ the optical depths differ by nearly a factor of two, suggesting that future high-redshift $\tau$ measurements could constrain the intensity of an early X-ray background.

\subsection{Caveats}
\label{sec:caveat}

As outlined in Section~\ref{sec:pop3}, this work adopts an idealised treatment of Pop~III stars by assuming a fixed mass, metal yield, and evolutionary pathway. Consequently, the impact of the Pop~III IMF on the early X-ray background remains an open question. According to the stellar evolutionary model of \citet{heger_nucleosynthetic_2002}, a $100~\msun$ Pop~III star collapses directly into a black hole and may subsequently produce hard X-ray photons through gas accretion. Varying Pop~III mass would therefore modulate the relative strengths of soft and hard X-ray photons. Previous studies have found that hard X-ray photons are less effective at promoting Pop~III star formation \citep{machacek_effects_2003,jeon_radiative_2014,hummel_first_2015}. Therefore, the positive X-ray feedback reported here could be weaker if a significant fraction of Pop~III stars collapse into accreting black holes. In addition, lower mass Pop~III stars explode as less energetic hypernovae \citep{nomoto_nucleosynthesis_2006}, which would further reduce the X-ray energy. Another interesting implication is that estimating the relative strength of soft and hard X-ray backgrounds in the early Universe may provide an indirect constraint on Pop~III IMF, highlighting the need for future studies that explicitly model X-ray sources associated with Pop~III stars with different masses.

\citet{skinner_cradles_2020} and \citet{wells_connecting_2022} allowed multiple Pop~III stars to form within a halo. In contrast, we limit the number of Pop~III star formation event to one per halo, while assuming a Pop~III multiplicity within that event. Adopting a relatively high multiplicity ($N \sim 10$, Papar~II; \citealp{prole_fragmentation-induced_2022}; Paper~III) and assuming that all Pop~III stars explode as PISNe, we use a large supernova boost factor $\alpha_{HN} = 1000$.\footnote{It is thought that Pop~III stars radiate at the Eddington rates, $L \sim 10^{38}(M/\msun)$~erg~s$^{-1}$ \citep{bromm_generic_2001}, and therefore LW luminosity of a $100~\msun$ Pop~III particle is insensitive to the assumed multiplicity.} Given the uncertainties in Pop~III multiplicity \citep{prole_fragmentation-induced_2022} and non-linear relationship between Pop~III mass and evolutionary pathways, it is difficult to predict their impact on the X-ray background production. Nevertheless, we expect that adopting a mass-dependent supernova model \citep{heger_nucleosynthetic_2002} would reduce the X-ray feedback produced by Pop~III supernovae in general. Another simplification in this work is that the Pop~III is assumed to be independent of the X-ray background. Our previous works (Papers~I--III) found that the characteristic Pop~III mass and multiplicity decrease in stronger X-ray backgrounds. Incorporating this dependence self-consistently into the X-ray background calculation would therefore reduce the global X-ray background as it builds up over time.

In this work, we optimistically assume that all Pop~III stars explode as PISNe and adopt an enhanced X-ray boost factor. Although values more consistent with stellar evolution models \citep[e.g.,][]{heger_nucleosynthetic_2002} will be explored in future work, we can assess the impact of this assumption on one of our main conclusions: the effectiveness of the on-the-fly method. In Section~\ref{sec:compare} we compare the global X-ray intensities derived from the two methods: on-the-fly (OTF) and post-processing (PP). The two approaches yield X-ray intensities of J$_{OTF} \sim 10^{-5}$ and J$_{PP} \sim 10^{-7}$, resulting in 7 and 3 Pop~III stars, respectively. Changing the boost factor ($\alpha_{HN}$) would alter the absolute level of X-ray feedback and therefore the expected number of Pop~III stars, but the qualitative conclusion remains unchanged provided the modification is moderate. For example, adopting $\alpha_{HN} = 10$ would reduce the intensities to J$_{OTF} \sim 10^{-7}$ and J$_{PP} \sim 10^{-9}$, corresponding to approximately 3 and 1 Pop~III stars respectively. Thus, the on-the-fly method would still predict a stronger X-ray feedback than the post processing approach. If a smaller value ($\alpha_{HN} = 1$) were adopted, neither method would produce significant positive X-ray feedback from Pop~III supernovae alone, and other X-ray sources such as HMXBs could become important.

\subsection{Summary}
\label{sec:summary}

We have investigated how an X-ray background in the early Universe influences primordial gas, computing the radiation background on-the-fly self-consistently with the simulations' star formation history to capture the feedback loop between star formation and the evolving background. The key results are as follows.
\begin{enumerate}
    \item When Pop~III supernovae explode as PISNe, they produce a weak ($\intj_{21} \sim 10^{-5}$) soft (peaked at $E \sim 300$~eV) X-ray background at $z \sim 15$. A moderate LW background ($\intj_{21} \sim 10^{-1}$) is established above $z \sim 12$ and strengthens rapidly below this redshift with the onset of Pop~II star formation.

    \item The X-ray background modifies the global properties of primordial gas through heating and ionisation. The increase in the \hm fraction (equation~(\ref{eq:hminus})) is less pronounced, because the LW background efficiently photodissociates \hm in the diffuse IGM. However, gas within minihaloes retains a higher \hm fraction under X-ray irradiation, promoting Pop~III star formation as shown in Fig.~\ref{fig:npop3}.

    \item The on-the-fly X-ray background is $1$--$2$ orders of magnitude stronger than the post-processed estimate, because it captures the reinforcing interplay between Pop~III star formation and the radiation backgrounds they produce. When applied to the same initial conditions, the on-the-fly approach yields significantly more Pop~III stars (Fig.~\ref{fig:post}).

    \item The X-ray background partially ionises primordial gas and raises the Thomson scattering optical depth at $z \gtrsim z_{re} = 7.68$. The total optical depths in both cases remain consistent with Planck~2018 results within observational uncertainties.
\end{enumerate}

This paper focuses on the methodology for computing radiation backgrounds and their effects on the IGM. The implications for Pop~III star formation and number density, and future extensions of this work, are discussed in the companion paper PR26b.

Our X-ray background includes only Pop~III supernovae as sources. \textit{JWST} observations, however, point to a large population of black holes in the early Universe \citep{juodzbalis_epochs_2023, matthee_little_2024, fujimoto_uncover_2024, rusakov_jwsts_2025, kido_black_2025}. While the companion paper (PR26b, Appendix~A) presents test results with enhanced X-ray intensities, modelling X-ray emission from other sources (X-ray binaries and accreting black holes) is left for future work.

\section*{Acknowledgements}

We acknowledge the anonymous referee for constructive comments and helping us improve the quality and clarity of the paper. JP acknowledges the support of the NRF of Korea grant funded by the Korean government (MSIT, RS-2022-NR068800). This work was supported by the BK21 Fostering Outstanding Universities for Research (FOUR) Program (4120200513819). The authors acknowledge the University of Maryland supercomputing resources (\url{http://hpcc.umd.edu}) made available for conducting this research. This work was granted access to the HPC resources of KISTI under allocation KSC-2024-CRE-0200. Large data transfers were supported by KREONET, managed and operated by KISTI.

\section*{Data Availability}

The data underlying this article will be shared on reasonable request to the corresponding author.



\bibliographystyle{mnras}
\bibliography{reference} 






\bsp	
\label{lastpage}
\end{document}